\documentclass[11pt,a4paper]{article}
\pdfoutput=1  
\usepackage{epsfig}
\usepackage{graphicx,psfrag}
\usepackage{multirow}
\usepackage{slashed}
\usepackage{pstricks}
\usepackage{caption}
\usepackage{subcaption}
\usepackage{url}
\usepackage{braket}
\usepackage{jheppub}
\usepackage{enumerate}
\usepackage{wasysym} 
\usepackage{mathrsfs} 
\usepackage{amsfonts} 
\usepackage{amsbsy} 
\usepackage{amscd}
\usepackage{color}
\usepackage{changepage}
\usepackage{floatrow}

\makeatletter

\def\eeq{\end{equation}}
\def\beq{\begin{equation}}

\newcommand{\Rmnum}[1]{\expandafter\@slowromancap\romannumeral #1@}

\newcommand{\bea} {\begin{eqnarray}}
\newcommand{\eea} {\end{eqnarray}}

\newcommand{\lsim}{\raisebox{-0.13cm}{~\shortstack{$<$ \\[-0.07cm]
      $\sim$}}~}
\makeatother

\title{Vacuum (in)stability in 2HDMS vs N2HDM}
\author[a]{Jayita Lahiri,}
  
\author[a,b]{Gudrid Moortgat-Pick}

\affiliation[a]{II. Institut f{\"u}r Theoretische Physik, Universit{\"a}t Hamburg, Luruper Chaussee 149, 22761 Hamburg, Germany} 

\affiliation[b]{Deutsches Elektronen-Synchrotron DESY, Notkestr. 85, 22607 Hamburg, Germany}

\emailAdd{jayita.lahiri@desy.de}
\emailAdd{gudrid.moortgat-pick@desy.de}

\abstract{In this work, we examine the criteria for vacuum stability in two models with extended scalar sectors namely, the N2HDM and the 2HDMS and make a detailed comparison between the two. For the purpose of demonstration, we choose a scenario which can accommodate the recently observed 95 GeV excess in both models. We further explore the impact of the measurement of the Yukawa couplings, the gauge boson couplings and most importantly the trilinear self-couplings of the scalars, in distinguishing the vacuum structure in both models. We further investigate the constraints from vacuum stability on the 2HDMS scenario that accommodates a viable dark matter candidate and compare it with the N2HDM case.

}

\preprint{DESY-24-126}
\begin{document}

\maketitle

\newpage

\section{Introduction}
\label{sec1}

The Higgs mechanism in Standard Model~\cite{Higgs:1964ia,Higgs:1964pj,Englert:1964et,Guralnik:1964eu,Higgs:1966ev} provides an explanation for the masses of all SM fermions and gauge bosons, but relies on a stable electro-weak (EW) vacuum, which is guaranteed at tree-level in SM Higgs potential. However, in the presence of higher-order corrections the stability of the SM scalar potential is not obvious anymore and it depends strongly on the SM parameters, in particular on the top Yukawa coupling~\cite{Degrassi:2012ry,Bednyakov:2015sca}. However, when extended scalar sectors are appended into the SM, the vacuum configuration can change. In the presence of extra scalars the absolute stability of the EW vacuum is no longer guaranteed. On the other hand, the loop contributions of these additional scalars can counteract with loop contribution involving top Yukawa coupling and have further impact the stability of the EW vacuum further. In the presence of such extra scalar degrees of freedom, there can exist charge or CP-breaking vacua simultaneously with the correct EW vacuum, so that the conservation of electric charge or CP-symmetry gets broken. Furthermore, there can also be simultaneous EW vacua which do not correspond to the correct EW vacuum expectation value (vev), i.e. $v\neq246$ GeV. Such type of vacuum is called {\it panic vacuum}~\cite{Ivanov:2006yq,Ivanov:2007de,Barroso:2007rr,Barroso:2012mj,Barroso:2013awa}. 

 If either charge/CP-breaking or {\it panic} vacua exist that are deeper than the EW vacuum, and the transition time from EW vacuum to the additional vacua is larger than the age of the universe, in that case the EW vacuum is considered to be metastable. However, there can be regions of the parameter space for specific models where the transition time to such dangerous vacua is less than the age of the universe. Such scenarios are considered to be unphysical and the corresponding parameter space can be excluded from the stability point of view. 

 The stability criteria in the context of the 2HDM have been studied in detail in \cite{Ivanov:2006yq,Ivanov:2007de,Ferreira:2004yd}. It has been shown that in the 2HDM, all the charge or CP-breaking vacua are necessarily saddle points and that they exist above the `normal' EW vacuum. Therefore, stability with respect to charge and CP-breaking is guaranteed in 2HDM. However, the wrong EW vacuum or {\it panic} vacuum can still exist below the EW minimum and the parameter space should be checked with respect to such kind of dangerous vacua. 
 The vacuum stability of the 2HDM at finite temperature and the phenomenon of {\it vacuum trapping} have been studied in~\cite{Biekotter:2021ysx,Biekotter:2022kgf}.
 The vacuum structure of the Inert Doublet model is discussed in~\cite{Ferreira:2015pfi} including the effect of loop contributions. Furthermore, the strong first order phase transition in case of both CP-conserving~\cite{Basler:2016obg} and CP-violating~\cite{Basler:2017uxn} 2HDM has also been studied in detail. 
 The vacuum stability in the supersymmetric models has been explored in~\cite{Hollik:2018wrr,Biekotter:2021rak}.

 The real scalar extension of the 2HDM, namely the N2HDM has been studied in the context of vacuum stability at zero temperature in detail in~\cite{Ferreira:2019iqb}. Furthermore, the stability of the scalar potential in N2HDM at finite temperatures and the phenomenon of {\it vacuum trapping} has also been explored~\cite{Biekotter:2021ysx}. It has been shown that the vacuum stability of the 2HDM gets weaker in the presence of an additional real singlet. The corresponding part of the parameter space of N2HDM has also been identified which is absolutely unstable and therefore unphysical. 

 At this point, it is only natural to ask, how the vacuum stability is affected when we extend our studies to a complex singlet extension of the 2HDM, the 2HDMS~\cite{Heinemeyer:2021msz,Dutta:2022xbd,Dutta:2023cig}, which involves an additional degree of freedom compared to the N2HDM. One of the crucial features of N2HDM and 2HDMS is that under an unbroken $Z_2$ symmetry, imposed on the scalar singlet, the model can also accommodate a dark matter (DM) candidate. If the $Z_2$-symmetry is broken by a singlet vev, both models can accommodate the recently observed 95 GeV excess at the CMS~\cite{CMS-PAS-HIG-20-002} and ATLAS~\cite{ATLAS-CONF-2018-025} experiments. Interestingly, the complex singlet extension namely, 2HDMS can accommodate both a DM candidate as well as the recent 95 GeV excess simultaneously, with a certain symmetry-breaking pattern in the scalar sector, and we call this scenario `dark 2HDMS'. It is also noteworthy that, under different symmetry conditions imposed on the additional scalar sector, the 2HDMS can lead to substantially different phenomenology. In this work, we have considered the $Z_2$ symmetry on the additional complex scalar. Other variants of this model, e.g, the 2HDMS with a $Z_3$-symmetry in~\cite{Heinemeyer:2021msz} and $U(1)$-symmetry in S2HDM~\cite{Biekotter:2021ovi} exist in the literature.

 In this study, we explore the vacuum structure of the 2HDMS with an imposed $Z_2$-symmetry on the complex singlet and perform a comparison between 2HDMS and N2HDM in this regard. First, a general analytical analysis is done and numerical results in terms of a chosen benchmark are presented. We impose observational constraints on the two models, and study whether the vacuum stability in the two models hints at some differences between the two models. We explore the role of Yukawa and gauge boson couplings of the scalars and, most importantly, the trilinear coupling of the Higgs boson. We investigate the vacuum stability of the `dark 2HDMS'. The numerical analyses in earlier works have been performed with codes such as $\tt BSMPT$~\cite{Basler:2018cwe} or {\tt Vivacious}~\cite{Camargo-Molina:2013qva,Camargo-Molina:2014pwa}. In this work, throughout the whole analysis, we use {\tt EVADE}~\cite{Hollik:2018wrr,Ferreira:2019iqb}. We also mention that our analysis is done at the leading order and assuming $T=0$.

 The plan of our work is as follows. In Section~\ref{sec2}, we discuss the two models under comparison, namely N2HDM and 2HDMS. In Section~\ref{sec3}, we present the analytical analysis. In Section ~\ref{sec4}, we discuss our numerical results, in terms of comparison between N2HDM and 2HDMS (without dark matter) for a chosen benchmark. In Section~\ref{sec5}, we concentrate in particular on the Yukawa, gauge boson couplings of the scalars and the trilinear Higgs coupling in both models. In Section~\ref{sec6}, we study our numerical results in comparison between N2HDM and dark 2HDMS. A short summary is given in Section~\ref{sec7}.

\section{The Models}
\label{sec2}
Our goal is to compare the vacuum structures of the two models, N2HDM~\cite{Cai:2013zga} and 2HDMS~\cite{Baum:2018zhf}, both involving extended scalar sectors. The vacuum (in)stability of N2HDM has already been studied in detail in~\cite{Ferreira:2019iqb}. We therefore ask, how much does the vacuum structure change when one extra degree of freedom is added to the N2HDM?

We briefly introduce the two models. Both N2HDM and 2HDMS contain the scalar potential ${\cal V}_{2HDM}$ of the 2HDM(Two-Higgs doublet model)~\cite{Branco:2011iw}, which is given as follows.

\begin{eqnarray} \label{V2HDM} {\cal V}_{2HDM} &=& m_{11}^2
(\Phi_1^{\dagger} \Phi_1) + m_{22}^2 (\Phi_2^{\dagger}
\Phi_2) - \left[m_{12}^2 (\Phi_1^{\dagger} \Phi_2 + \rm h.c.)\right]+ \frac{\lambda_1}{2}  (\Phi_1^{\dagger} \Phi_1)^2 +
\frac{\lambda_2}{2} (\Phi_2^{\dagger} \Phi_2)^2\nonumber \\
&&  + \lambda_3
(\Phi_1^{\dagger} \Phi_1)(\Phi_2^{\dagger} \Phi_2) + \lambda_4
(\Phi_1^{\dagger}
\Phi_2)(\Phi_2^{\dagger} \Phi_1)+ \left[\frac{\lambda_5}{2} (\Phi_1^{\dagger} \Phi_2)^2 + \rm
h.c.\right].
\end{eqnarray}

\noindent
It is worth mentioning that, in order to avoid the dangerous flavor-changing-neutral currents(FCNC's) at the tree-level, a $Z_2$ symmetry is imposed on the 2HDM Lagrangian. In addition, we assume CP-conservation in the theory and therefore, all the coefficients in the potential~Eq.~(\ref{V2HDM}) are taken to be real.

The N2HDM, furthermore, contains an extra real scalar $S$ giving rise to additional terms in the scalar potential, including an imposed $Z_2'$ symmetry is imposed on the real scalar singlet field $S$.

\begin{equation}
{\cal V_{\text{N2HDM}}} = {\cal V}_{2HDM} + {\cal V}_S
\end{equation}

with

\begin{equation}
{\cal V}_S = \frac{1}{2} m_S^2 S^2 + \frac{1}{8}\lambda_6 S^4 + \frac{1}{2} \lambda_7 |\Phi_1|^2 S^2 + \frac{1}{2}\lambda_8  |\Phi_2|^2 S^2.
\label{n2hdm_lag}
\end{equation}
 
\noindent
On the other hand, in the 2HDMS with a complex singlet denoted here, as $S+iP$\footnote{We have named the real part of the complex scalar $S$ in order to make a direct comparison with the N2HDM potential}, there are additional terms in the scalar potential in terms of $V_S'$. Here too, a $Z_2'$ symmetry is imposed on the complex field $S+iP$. We mention one caveat here. Once we demand that all the coefficients in the Lagrangian have to be real, all the terms involving odd powers of $S$ or $P$ get canceled and we are left with the following form of the potential. 

\begin{equation}
{\cal V_{\text{2HDMS}}} = {\cal V}_{2HDM} + {\cal V}_S',
\end{equation}

\noindent
where
\begin{eqnarray}
{\cal V}_S' = \frac{1}{2} m_S^2 S^2 + \frac{1}{2}m_{S'}^2 P^2 + \frac{1}{8}\lambda_6 S^4 + \frac{1}{8} \lambda_9 P^4 + \frac{1}{8} \lambda_{10} S^2 P^2 \nonumber \\ + \frac{1}{2} (\lambda_7 |\Phi_1|^2 + \lambda_8  |\Phi_2|^2) S^2 + \frac{1}{2} (\lambda_{11} |\Phi_1|^2 + \lambda_{12} |\Phi_2|^2) P^2.
\label{2hdms_lag}
\end{eqnarray}

\noindent
It is clear from Eq.~(\ref{2hdms_lag}), that this potential looks very similar to a scalar potential involving two real scalars transforming under two different $Z_2$ symmetries. We explored this idea and have added a discussion on the similarities between the complex singlet and two real scalar singlets in Appendix~\ref{tworealscalar}.

 It is also worth mentioning that, a variant of this 2HDMS model has been studied in~\cite{Heinemeyer:2021msz}, where instead of a $Z_2'$ symmetry, a $Z_3$ symmetry is imposed on the complex singlet. These two models are fundamentally very different due to the different underlying symmetries and will lead to different phenomenologies as well. We have added a discussion on the differences between the $Z_2'$- and $Z_3$-symmetric 2HDMS models in the Appendix~\ref{z32hdms_compare}.

As mentioned before, we assume a $Z_2 \times Z_2'$ symmetry on the scalar potential in both cases, N2HDM and 2HDMS. We also assume all the parameters of our potential to be real. For clarity, we present the symmetry structure of both models in Table~\ref{table1}.

\begin{table}[!hptb]
\begin{center}
\begin{tabular}{| c | c | c |}
\hline
 & $Z_2$ & $Z_2'$  \\
\hline
$\Phi_1$ & + & + \\
\hline
$\Phi_2$ & - & + \\
\hline
$S$ & + & - \\
\hline
\end{tabular}
\quad
\begin{tabular}{| c | c | c |}
\hline
 & $Z_2$ & $Z_2'$  \\
\hline
$\Phi_1$ & + & + \\
\hline
$\Phi_2$ & - & + \\
\hline
$S$ and $P$  & + & - \\
\hline
\end{tabular}
\caption{\it Transformation of fields under $Z_2 \times Z_2'$ in N2HDM (left) as well as 2HDMS (right).}
\label{table1}
\end{center}
\end{table} 

\noindent
We can see from Eq.~(\ref{n2hdm_lag}) and (\ref{2hdms_lag}) that, compared to the N2HDM, the scalar potential of the 2HDMS is richer. Our goal is to study the impact of the extra degree of freedom as well as the new quartic couplings on the vacuum stabilities of the 2HDMS. Before comparing the vacuum structures of both models, we discuss the free parameters of the two models in the interaction basis as well as the mass basis. 

\begin{table}[!hptb]
\begin{flushleft}
\begin{tabular}{|c|c|c|}
\hline
 &  Interaction basis & Mass basis \\
\hline
N2HDM & $\lambda_{1,..8}, v_s, m_{12}^2$,$\tan\beta$, $v$ &  $m_{h_{1,..3}}, m_A, m_{H^{\pm}}$, $\alpha_{1,..3}$, $v_s$, $m_{12}^2$, $\tan\beta$, $v$\\
\hline
2HDMS($v_p=0$) & $\lambda_{1,..12}, v_s, m_S', m_{12}^2$,$\tan\beta$, $v$ & $m_{h_{1,..3}}, m_A, m_{H^{\pm}}$, $\alpha_{1,..3}$, $v_s$, $m_{DM}$, $\lambda_{9,..12}$, $m_{12}^2$, $\tan\beta$, $v$\\
\hline
2HDMS($v_p\neq0$) & $\lambda_{1,..12}, v_s, v_p, m_{12}^2$,$\tan\beta$, $v$ & $m_{h_{1,..4}}, m_A, m_{H^{\pm}}$, $\alpha_{1,..6}$, $v_s$, $v_p$, $m_{12}^2$, $\tan\beta$, $v$\\
\hline
\end{tabular}
\end{flushleft}
\caption{Independent parameters in interaction basis and mass basis for N2HDM and 2HDMS.}
\label{novev}
\end{table}

\noindent
The fields in the interaction basis can be transformed into the mass basis by a unitary transformation. 
In case of the N2HDM as well as the 2HDMS where P has no vev, the transformation can be done as follows.

\begin{equation}
\left(\begin{array}{c} \sqrt{2} Re \Phi_1^0 - v_1 \\
\sqrt{2} Re \Phi_2^0 - v_2 \\
S-v_s
\end{array}\right)\, 
=R \left(\begin{array}{c} h_1 \\
h_2 \\
h_3
\end{array}\right),
\label{n2hdm_basis}
\end{equation}

The transformation between interaction basis and mass basis in the case of the 2HDMS where $v_p \neq 0$, is given as follows.

\begin{equation}
\left(\begin{array}{c} \sqrt{2} Re \Phi_1^0 - v_1 \\
\sqrt{2} Re \Phi_2^0 - v_2 \\
S-v_s \\
P-v_p
\end{array}\right)\, 
=R' \left(\begin{array}{c} h_1 \\
h_2 \\
h_3 \\
h_4
\end{array}\right),
\label{2hdms_basis}
\end{equation}

\noindent
It can clearly be seen from Table~\ref{novev} that compared to the N2HDM, in the 2HDMS(vev of $P$ = 0 case) there are the four additional quartic couplings ($\lambda_9,\lambda_{10},\lambda_{11},\lambda_{12}$) and the additional mass parameter $m_S'$. Since in this case $P$ does not acquire vev and therefore does not mix with other scalar states of the model, owing to the $Z_2'$ symmetry, $P$ will act as a DM candidate and we call it the dark 2HDMS scenario. In this case, in the mass basis the extra parameters can be translated into the mass of the DM candidate ($m_{DM}$), the DM self-coupling $\lambda_9$ and DM portal couplings $\lambda_{10,..12}$ with scalars $h_{1,2,3}$. The expression for the dark matter mass ($m_{DM}$) is given as follows.

\begin{equation}
m_{DM}^2 = m_{S'}^2 + \frac{1}{4} \lambda_{10} v_s^2 + \frac{1}{2} (\lambda_{11} v_1^2 + \lambda_{12} v_2^2)
\label{dmmass}
\end{equation}

In the latter scenario, i.e, with non-zero vev for $P$, the additional parameters in the 2HDMS compared to the N2HDM, in the interaction basis there are the four additional quartic couplings ($\lambda_9,\lambda_{10},\lambda_{11},\lambda_{12}$)
and the vev ($v_p$) of the additional degree of freedom ($P$). Since in this case, both $S$ and $P$ acquire a  vev, both these degrees of freedom mix with the scalar degrees of freedom of the 2HDM, and we get in total four scalars in the theory. This mixing is a result of the $Z_2'$ symmetry imposed on the complex singlet(see Appendix~\ref{z32hdms_compare} for more details). Compared to N2HDM, the additional parameters in the 2HDMS correspond to one mass of the extra scalar $h_4$, three extra mixing angles ($\alpha_4, \alpha_5$ and $\alpha_6$), and the vev for $P$ ($v_p$) in the mass basis.

\section{Comparison between vacuum stability of N2HDM and 2HDMS : Analytical approach}
\label{sec3}

Having discussed the scalar potential, imposed symmetries and free parameters of the two models, we now explore the conditions of the stability of the EW vacuum in both models. In general terms, if there is no stationary point deeper than
the EW vacuum, then the EW vacuum is {\it absolutely stable}. If there are deeper minima (which can be charge, CP-breaking or neutral), we calculate the tunelling time from the EW vacuum to each of those deeper minima. The decay width per volume for tunelling into a deeper minima in field space is given as $K e^{-B}$~\cite{Coleman:1977py,Callan:1977pt}, where $B$ is called the bounce action. $K$ is a sub-dominant contribution which requires involved calculations and therefore is estimated on dimensional grounds~\cite{Adams:1993zs,Ferreira:2019iqb}. If $B > 390$, the transition time into deeper minima, is larger than the age of the universe and EW vacuum is called {\it metastable}, else it is called {\it unstable}~\cite{Ferreira:2019iqb}~\footnote{It is a conservative approach where the survival probability through the age of the universe $\lsim 5.73\times 10^{-7}$ are considered {\it short-lived} or {\it unstable}.}. The goal of this section is to derive conditions to identify such unstable (unphysical) vacua and to exclude such regions from the corresponding parameter space in both models.

\subsection{Possible charge and CP-breaking vacuua in N2HDM and 2HDMS}

Let us now compare the the conditions for charge-breaking vacua in N2HDM and 2HDMS. In order to study the interplay between
different possible vacua, it is convenient to adopt a bilinear formalism, which was first developed for
the 2HDM~\cite{Ferreira:2004yd} and then was extended and employed for the N2HDM~\cite{Ferreira:2019iqb}. In this work we further extend the formalism for 2HDMS. The relevant bilinears in this case will be the following.

\begin{eqnarray*}
x_1 = |\Phi_1|^2, x_2 = |\Phi_2|^2, x_3 = \text{Re}(\Phi_1^{\dagger}\Phi_2) 
, x_4 = \text{Im}(\Phi_1^{\dagger}\Phi_2), x_5 = |S|^2, x_6= |P|^2.
\end{eqnarray*}

\noindent
In the N2HDM, there are two possible EW vacua, namely ${\cal N}$-type vacuum (EW vacuum with no singlet vev i.e. $v_s=0$) and ${\cal N}_s$-type vacuum (EW vacuum with scalar non-zero singlet vev $v_s$).  

\begin{equation}
{\cal N} \rightarrow
\braket{\Phi_1}_0=\frac{1}{\sqrt{2}}\left(\begin{array}{c} 0 \\
\,v_1
\end{array}\right)\,, \ \ \
\braket{\Phi_2}_0=\frac{1}{\sqrt{2}}\left(\begin{array}{c} 0 \\
\,v_2
\end{array}\right)\,, \ \ \
\braket{S}_0=0,
\end{equation}

\begin{equation}
{\cal N}_s \rightarrow
\braket{\Phi_1}_0=\frac{1}{\sqrt{2}}\left(\begin{array}{c} 0 \\
\,v_1
\end{array}\right)\,, \ \ \
\braket{\Phi_2}_0=\frac{1}{\sqrt{2}}\left(\begin{array}{c} 0 \\
\,v_2
\end{array}\right)\,, \ \ \
\braket{S}_0=v_s.
\end{equation}

\noindent
The possible charge-breaking vacua of the N2HDM are ${\cal CB}$ and ${\cal CB}_s$. They can be written in the following forms.

 \begin{equation}
{\cal CB} \rightarrow
\braket{\Phi_1}_0=\frac{1}{\sqrt{2}}\left(\begin{array}{c} 0 \\
\,c_1
\end{array}\right)\,, \ \ \
\braket{\Phi_2}_0=\frac{1}{\sqrt{2}}\left(\begin{array}{c} c_2 \\
\,c_3
\end{array}\right)\,, \ \ \
\braket{S}_0=0,
\end{equation}

\begin{equation}
{\cal CB}_s \rightarrow
\braket{\Phi_1}_0=\frac{1}{\sqrt{2}}\left(\begin{array}{c} 0 \\
\,c_1
\end{array}\right)\,, \ \ \
\braket{\Phi_2}_0=\frac{1}{\sqrt{2}}\left(\begin{array}{c} c_2 \\
\,c_3
\end{array}\right)\,, \ \ \
\braket{S}_0=c_4.
\end{equation}

\noindent
The possible ${\cal CP}$-breaking vacua of N2HDM are ${\cal CP}$ and ${\cal CP}_s$. They can be written in the following forms.

 \begin{equation}
{\cal CP} \rightarrow
\braket{\Phi_1}_0=\frac{1}{\sqrt{2}}\left(\begin{array}{c} 0 \\
\,c_1'
\end{array}\right)\,, \ \ \
\braket{\Phi_2}_0=\frac{1}{\sqrt{2}}\left(\begin{array}{c} 0 \\
\,c_2' + i c_3'
\end{array}\right)\,, \ \ \
\braket{S}_0=0,
\end{equation}

\begin{equation}
{\cal CP}_s \rightarrow
\braket{\Phi_1}_0=\frac{1}{\sqrt{2}}\left(\begin{array}{c} 0 \\
\,c_1'
\end{array}\right)\,, \ \ \
\braket{\Phi_2}_0=\frac{1}{\sqrt{2}}\left(\begin{array}{c} 0 \\
\,c_2' + i c_3'
\end{array}\right)\,, \ \ \
\braket{S}_0=c_4'.
\end{equation}

\noindent
Compared to the N2HDM, the additional degree of freedom $P$ of 2HDMS can also acquire a vev. Therefore, in the 2HDMS, there are four possible EW vacua.

\begin{equation}
{\cal N} \rightarrow
\braket{\Phi_1}_0=\frac{1}{\sqrt{2}}\left(\begin{array}{c} 0 \\
\,v_1
\end{array}\right)\,, \ \ \
\braket{\Phi_2}_0=\frac{1}{\sqrt{2}}\left(\begin{array}{c} 0 \\
\,v_2
\end{array}\right)\,, \ \ \
\braket{S}_0=0, \ \ \ \braket{P}_0=0,
\end{equation}

\begin{equation}
{\cal N}_{s} \rightarrow
\braket{\Phi_1}_0=\frac{1}{\sqrt{2}}\left(\begin{array}{c} 0 \\
\,v_1
\end{array}\right)\,, \ \ \
\braket{\Phi_2}_0=\frac{1}{\sqrt{2}}\left(\begin{array}{c} 0 \\
\,v_2
\end{array}\right)\,, \ \ \
\braket{S}_0=v_s, \ \ \ \braket{P}_0=0,
\end{equation}

\begin{equation}
{\cal N}_p \rightarrow
\braket{\Phi_1}_0=\frac{1}{\sqrt{2}}\left(\begin{array}{c} 0 \\
\,v_1
\end{array}\right)\,, \ \ \
\braket{\Phi_2}_0=\frac{1}{\sqrt{2}}\left(\begin{array}{c} 0 \\
\,v_2
\end{array}\right)\,, \ \ \
\braket{S}_0=0, \ \ \ \braket{P}_0=v_p,
\end{equation}

\begin{equation}
{\cal N}_{sp} \rightarrow
\braket{\Phi_1}_0=\frac{1}{\sqrt{2}}\left(\begin{array}{c} 0 \\
\,v_1
\end{array}\right)\,, \ \ \
\braket{\Phi_2}_0=\frac{1}{\sqrt{2}}\left(\begin{array}{c} 0 \\
\,v_2
\end{array}\right)\,, \ \ \
\braket{S}_0=v_s, \ \ \ \braket{P}_0=v_p.
\end{equation}

\noindent
The possible charge-breaking vacua of 2HDMS are ${\cal CB}$, ${\cal CB}_s$, ${\cal CB}_p$ and ${\cal CB}_{sp}$. They can be written in the following forms.

\begin{equation}
{\cal CB} \rightarrow
\braket{\Phi_1}_0=\frac{1}{\sqrt{2}}\left(\begin{array}{c} 0 \\
\,c_1
\end{array}\right)\,, \ \ \
\braket{\Phi_2}_0=\frac{1}{\sqrt{2}}\left(\begin{array}{c} c_2 \\
\,c_3
\end{array}\right)\,, \ \ \
\braket{S}_0=0,\ \ \ \braket{P}_0=0,
\end{equation}

\begin{equation}
{\cal CB}_s \rightarrow
\braket{\Phi_1}_0=\frac{1}{\sqrt{2}}\left(\begin{array}{c} 0 \\
\,c_1
\end{array}\right)\,, \ \ \
\braket{\Phi_2}_0=\frac{1}{\sqrt{2}}\left(\begin{array}{c} c_2 \\
\,c_3
\end{array}\right)\,, \ \ \
\braket{S}_0=c_4,\ \ \ \braket{P}_0=0,
\end{equation}

\begin{equation}
{\cal CB}_p \rightarrow
\braket{\Phi_1}_0=\frac{1}{\sqrt{2}}\left(\begin{array}{c} 0 \\
\,c_1
\end{array}\right)\,, \ \ \
\braket{\Phi_2}_0=\frac{1}{\sqrt{2}}\left(\begin{array}{c} c_2 \\
\,c_3
\end{array}\right)\,, \ \ \
\braket{S}_0=0,\ \ \ \braket{P}_0=c_5,
\end{equation}

\begin{equation}
{\cal CB}_{sp} \rightarrow
\braket{\Phi_1}_0=\frac{1}{\sqrt{2}}\left(\begin{array}{c} 0 \\
\,c_1
\end{array}\right)\,, \ \ \
\braket{\Phi_2}_0=\frac{1}{\sqrt{2}}\left(\begin{array}{c} c_2 \\
\,c_3
\end{array}\right)\,, \ \ \
\braket{S}_0= c_4, \ \ \ \braket{P}_0=c_5.
\end{equation}

\noindent
The possible ${\cal CP}$-breaking vacuua of 2HDMS are ${\cal CP}$, ${\cal CP}_s$, ${\cal CP}_p$ and ${\cal CP}_{sp}$. They can be written in the following forms.

\begin{equation}
{\cal CP} \rightarrow
\braket{\Phi_1}_0=\frac{1}{\sqrt{2}}\left(\begin{array}{c} 0 \\
\,c_1'
\end{array}\right)\,, \ \ \
\braket{\Phi_2}_0=\frac{1}{\sqrt{2}}\left(\begin{array}{c} 0 \\
\,c_2' + i c_3'
\end{array}\right)\,, \ \ \
\braket{S}_0=0,\ \ \ \braket{P}_0=0,
\end{equation}

\begin{equation}
{\cal CP}_s \rightarrow
\braket{\Phi_1}_0=\frac{1}{\sqrt{2}}\left(\begin{array}{c} 0 \\
\,c_1'
\end{array}\right)\,, \ \ \
\braket{\Phi_2}_0=\frac{1}{\sqrt{2}}\left(\begin{array}{c} 0 \\
\,c_2' + i c_3'
\end{array}\right)\,, \ \ \
\braket{S}_0=c_4',\ \ \ \braket{P}_0=0,
\end{equation}

\begin{equation}
{\cal CP}_p \rightarrow
\braket{\Phi_1}_0=\frac{1}{\sqrt{2}}\left(\begin{array}{c} 0 \\
\,c_1
\end{array}\right)\,, \ \ \
\braket{\Phi_2}_0=\frac{1}{\sqrt{2}}\left(\begin{array}{c} 0 \\
\,c_2' + i c_3'
\end{array}\right)\,, \ \ \
\braket{S}_0=0,\ \ \ \braket{P}_0=c_5',
\end{equation}

\begin{equation}
{\cal CP}_{sp} \rightarrow
\braket{\Phi_1}_0=\frac{1}{\sqrt{2}}\left(\begin{array}{c} 0 \\
\,c_1
\end{array}\right)\,, \ \ \
\braket{\Phi_2}_0=\frac{1}{\sqrt{2}}\left(\begin{array}{c} 0 \\
\,c_2' + i c_3'
\end{array}\right)\,, \ \ \
\braket{S}_0= c_4', \ \ \ \braket{P}_0=c_5'.
\end{equation}

\noindent
One can see that the number of possible unphysical or dangerous charge/CP-breaking vacua is larger in the 2HDMS compared to the N2HDM. Therefore, it is expected that the stability will get worse in the 2HDMS compared to the N2HDM. We will derive and discuss such results in the upcoming subsections. One example for the detailed calculation of the following results has been elaborated in detail in Appendix~\ref{calculation}.


\subsubsection{${\cal N}$-type vacuum}

The conditions for absolute stability of ${\cal N}$-type vacua are given by

\begin{eqnarray}
\label{eqnn5}
{\cal V}_{\cal CB} - {\cal V}_{\cal N} = (\frac{m_{H^{\pm}}^2}{2v^2})_{\cal N}[(v_2 c_1 - v_1 c_3)^2 + v_1^2 c_2^2] > 0,  \nonumber \\
{\cal V}_{{\cal CB}_{s}} - {\cal V}_{\cal N} = (\frac{m_{H^{\pm}}^2}{2v^2})_{\cal N}[(v_2 c_1 - v_1 c_3)^2 + v_1^2 c_2^2] + c_4^2 m_{s}^2 > 0, \nonumber \\ 
{\cal V}_{{\cal CB}_{p}} - {\cal V}_{\cal N} = (\frac{m_{H^{\pm}}^2}{2v^2})_{\cal N}[(v_2 c_1 - v_1 c_3)^2 + v_1^2 c_2^2] + c_5^2 m_{p}^2 > 0, \nonumber \\ 
{\cal V}_{{\cal CB}_{sp}} - {\cal V}_{\cal N} = (\frac{m_{H^{\pm}}^2}{2v^2})_{\cal N}[(v_2 c_1 - v_1 c_3)^2 + v_1^2 c_2^2] + c_4^2 m_{s}^2 + c_5^2 m_{p}^2  > 0.
\end{eqnarray} 

\noindent
The corresponding conditions for ${\cal CP}$-breaking breaking vacuua can be obtained by making the following transformation in Eq.~(\ref{eqnn5}). 

\begin{equation}
    c_1 \rightarrow c_1', c_3 \rightarrow c_2', c_2 \rightarrow c_3', c_4 \rightarrow c_4', c_5 \rightarrow c_5'.
\end{equation}

\noindent
First two equations in Eq.~(\ref{eqnn5}) are applicable to both N2HDM and 2HDMS and the last two equations pertain to 2HDMS. $m_s$(for both N2HDM and 2HDMS) and $m_p$(for 2HDMS), are the masses of the extra scalar degrees of freedom, which do not mix with the other scalars due to imposed $Z_2'$ symmetry and therefore can act as DM candidates in both models.  
It is evident that all of Eq.~(\ref{eqnn5}) are trivially satisfied at the ${\cal N}$-type minima in both N2HDM and 2HDMS, a result similar as in the 2HDM~\cite{Ferreira:2004yd}.


\subsubsection{${\cal N}_s$-type vacuum}

The condition for absolute stability of a ${\cal N}_s$-type vacuum, against a charge-breaking vacuum(${\cal CB}$) is given by the condition.

\begin{equation}
{\cal V}_{\cal CB} - {\cal V}_{{\cal N}_s} > 0
\label{eqn1}
\end{equation}

\noindent
Where
\begin{equation}
{\cal V}_{\cal CB} - {\cal V}_{{\cal N}_s} = (\frac{m_{H^{\pm}}^2}{2v^2})_{{\cal N}_s}[(v_2 c_1 - v_1 c_3)^2 + v_1^2 c_2^2] - v_s^2 m_{S1}^2
\label{eqn2}
\end{equation} 

\noindent
In Eq.~(\ref{eqn2}), $v_1$ and $v_2$ are the vev's of the two doublets (scalar components) in ${\cal N}_s$ scenario, and $c_1$ and $c_3$ are the vev's of the scalar components of the doublets and $c_3$ is the vev of the charged component of one of the doublets in the ${\cal CB}$ scenario and $v_s$ is the vev of the scalar part of the singlet.  

In both N2HDM and 2HDMS we derive,

\begin{equation}
\label{eqmass1}
m_{S1}^2 = m_S^2 + {\lambda_7}c_1^2 + {\lambda_8}(c_2^2 + c_3^2)
\end{equation}

\noindent
In the 2HDMS, in addition there is a possibility for ${\cal N}_s$-type vacua to transition into ${\cal CB}_p$-type ($P$ taking vev $c_5$) vacua. The stability condition for such a transition is 

\begin{equation}
{\cal V}_{{\cal CB}_{p}} - {\cal V}_{{\cal N}_s} > 0,
\label{eqnn11}
\end{equation}

\noindent
Where
\begin{equation}
{\cal V}_{{\cal CB}_{p}} - {\cal V}_{{\cal N}_s} = (\frac{m_{H^{\pm}}^2}{2v^2})_{{\cal N}_s}[(v_2 c_1 - v_1 c_3)^2 + v_1^2 c_2^2] + c_5^2 m_p^2 - v_s^2 m_{S1}^2
\label{eqnn3}
\end{equation}

\noindent
Where $m_p^2 = m_{S'}^2 + \frac{1}{4} \lambda_{10} v_s^2 + \frac{1}{2} (\lambda_{11} v_1^2 + \lambda_{12} v_2^2)$. One should note that ${{\cal N}_s}$-type vacuum corresponds to the dark 2HDMS and $m_p$ can be identified as the corresponding DM mass $m_{DM}$. On the other hand, $m_{S1}$ is given by,

\begin{equation}
\label{eqmass2}
m_{S1}^2 = m_S^2 + {\lambda_7}c_1^2 + {\lambda_8}(c_2^2 + c_3^2) + \frac{1}{4} \lambda_{10} c_5^2.
\end{equation}

\noindent
In a physical minima, $m_p^2$ is positive. However, $\lambda_{10}$ will play an important role in deciding whether, this condition is more stringent compared to the N2HDM. Therefore compared to N2HDM, 2HDMS with a scalar vev can have an additional type of charge-breaking minima.


\noindent
On the other hand,

\begin{equation}
{\cal V}_{{\cal CB}_{s}} - {\cal V}_{{\cal N}_s} = (\frac{m_{H^{\pm}}^2}{2v^2})_{{\cal N}_s}[(v_2 c_1 - v_1 c_3)^2 + v_1^2 c_2^2]. 
\label{eqnn4}
\end{equation} 
Therefore, the stability condition for ${\cal N}_s$-type vacua against ${\cal CB}_s$-type charge-breaking minima, i.e
${\cal V}_{{\cal CB}_{s}} - {\cal V}_{{\cal N}_s} > 0$, is trivially satisfied.

\noindent
Finally we get,

\begin{equation}
{\cal V}_{{\cal CB}_{sp}} - {\cal V}_{{\cal N}_s} = (\frac{m_{H^{\pm}}^2}{2v^2})_{{\cal N}_s}[(v_2 c_1 - v_1 c_3)^2 + v_1^2 c_2^2] + c_5^2 m_p^2.
\label{eqnnn5}
\end{equation}

\noindent
Considering the fact that $m_p^2$ is a physical mass-square in Eq.~(\ref{eqnnn5}) the condition ${\cal V}_{{\cal CB}_{sp}} - {\cal V}_{{\cal N}_s} > 0$ is trivially satisfied. Therefore, a ${\cal N}_s$ vacuum is also absolutely stable against ${\cal CB}_{sp}$-type charge-breaking minima.


\noindent
Here too, the corresponding conditions for ${\cal CP}$-breaking vacua can be obtained by applying the following transformation in Eqs.~(\ref{eqn2}), (\ref{eqmass1}),~(\ref{eqnn3})-(\ref{eqnnn5}). 

\begin{equation}
    c_1 \rightarrow c_1', c_3 \rightarrow c_2', c_2 \rightarrow c_3', c_4 \rightarrow c_4', c_5 \rightarrow c_5'.
\end{equation}

\subsubsection{${\cal N}_p$-type vacuum}

Let us now consider EW vacuum ${\cal N}_p$, where $v_s=0$ and $v_p\neq 0$. This vacuum only exists in the 2HDMS and not in the N2HDM.
As seen before, there are four possible charge-breaking minima ${\cal CB}$, ${\cal CB}_s$ ${\cal CB}_p$ and ${\cal CB}_{sp}$ . 

The stability condition of ${\cal N}_p$ against ${\cal CB}$ is given as follows,

\begin{equation}
{\cal V}_{\cal CB} - {\cal V}_{{\cal N}_p} > 0
\label{eqnn12}
\end{equation}

\noindent
where 
\begin{equation}
{\cal V}_{\cal CB} - {\cal V}_{{\cal N}_p} = (\frac{m_{H^{\pm}}^2}{2v^2})_{{\cal N}_p}[(v_2 c_1 - v_1 c_3)^2 + v_1^2 c_2^2] - m_{S2}^2 v_p^2,
\label{eqnn6}
\end{equation} 

\noindent
and

\begin{equation}
\label{masseq3}
m_{S2}^2 = m_S'^2 + {\lambda_{11}}c_1^2 + {\lambda_{12}}(c_2^2 + c_3^2).
\end{equation}

\noindent
On the other hand, the stability condition for ${\cal N}_p$ against ${\cal CB}_s$ is given by the following,

\begin{equation}
{\cal V}_{{\cal CB}_{s}} - {\cal V}_{{\cal N}_p} > 0
\label{eqnn13}
\end{equation}

\noindent
where, 

\begin{equation}
{\cal V}_{{\cal CB}_{s}} - {\cal V}_{{\cal N}_p} = (\frac{m_{H^{\pm}}^2}{2v^2})_{{\cal N}_p}[(v_2 c_1 - v_1 c_3)^2 + v_1^2 c_2^2] + c_4^2 m_s^2 - m_{S2}^2 v_p^2.
\label{eqnn7}
\end{equation}

\noindent
Here $m_s^2$ is the DM mass given by ($m_{S}^2 + \frac{1}{4} \lambda_{10} p^2 + \frac{1}{2} (\lambda_{7} v_1^2 + \lambda_{8} v_2^2)$) and 
 
\begin{equation}
\label{masseq4}
m_{S2}^2 = m_S'^2 + {\lambda_{11}}c_1^2 + {\lambda_{12}}(c_2^2 + c_3^2) + \frac{1}{4} \lambda_{10} c_4^2.
\end{equation}

\noindent
Now we derive, 

\begin{equation}
{\cal V}_{{\cal CB}_{p}} - {\cal V}_{{\cal N}_p} = (\frac{m_{H^{\pm}}^2}{2v^2})_{{\cal N}_p}[(v_2 c_1 - v_1 c_3)^2 + v_1^2 c_2^2]
\label{eqnnn8}
\end{equation}

\noindent
and ${\cal V}_{{\cal CB}_{p}} - {\cal V}_{{\cal N}_p} > 0$ is trivially satisfied. Therefore, ${\cal N}_p$ is absolutely stable against ${\cal CB}_p$-type charge-breaking minima.

Finally, we get

\begin{equation}
{\cal V}_{{\cal CB}_{sp}} - {\cal V}_{{\cal N}_p} = (\frac{m_{H^{\pm}}^2}{2v^2})_{{\cal N}_p}[(v_2 c_1 - v_1 c_3)^2 + v_1^2 c_2^2] +  c_4^2 m_s^2,
\label{eqnnn9}
\end{equation}

\noindent
$m_s^2$ being the physical DM mass in Eq.~(\ref{eqnnn9}), and the condition ${\cal V}_{{\cal CB}_{sp}} - {\cal V}_{{\cal N}_p} > 0$ is trivially satisfied too. Therefore, ${\cal N}_p$ is also absolutely stable against ${\cal CB}_{sp}$-type charge-breaking minima.

The corresponding conditions for ${\cal CP}$-breaking vacua can be obtained by applying the following transformation in Eqs.~(\ref{eqnn6}), (\ref{masseq3}), (\ref{eqnn7})-(\ref{eqnnn9}).

\begin{equation}
    c_1 \rightarrow c_1', c_3 \rightarrow c_2', c_2 \rightarrow c_3', c_4 \rightarrow c_4', c_5 \rightarrow c_5'
\end{equation}

\subsubsection{${\cal N}_{sp}$-type vacuum}

\noindent
We will now consider ${\cal N}_{sp}$-type EW vacua where both $S$ and $P$ fields acquire a vev. It is clear that this scenario can only exist in the 2HDMS.

First we will present the stability condition for such EW vacuum against ${\cal CB}$-type charge-breaking minima given:

\begin{equation}
{\cal V}_{\cal CB} - {\cal V}_{{\cal N}_{sp}} > 0,
\label{sp1}
\end{equation}

\noindent
where
\begin{equation}
{\cal V}_{\cal CB} - {\cal V}_{{\cal N}_{sp}} = (\frac{m_{H^{\pm}}^2}{2v^2})_{{\cal N}_{sp}}[(v_2 c_1 - v_1 c_3)^2 + v_1^2 c_2^2] - v_s^2 m_{S1}^2 -v_p^2 m_{S2}^2
\label{eqnn8}
\end{equation}

\noindent
and 
\begin{eqnarray}
\label{masseq5}
m_{S1}^2 = m_S^2 + {\lambda_7}c_1^2 + {\lambda_8}(c_2^2 + c_3^2), \nonumber \\
m_{S2}^2 = m_S'^2 + {\lambda_{11}}c_1^2 + {\lambda_{12}}(c_2^2 + c_3^2).
\end{eqnarray}

\noindent
Now for the stability against ${\cal CB}_s$ vacuum, one requires

\begin{equation}
{\cal V}_{{\cal CB}_{s}} - {\cal V}_{{\cal N}_{sp}} > 0,
\label{sp2}
\end{equation}

\noindent
Where
\begin{equation}
{\cal V}_{{\cal CB}_{s}} - {\cal V}_{{\cal N}_{sp}} = (\frac{m_{H^{\pm}}^2}{2v^2})_{{\cal N}_{sp}}[(v_2 c_1 - v_1 c_3)^2 + v_1^2 c_2^2] - v_p^2 m_{S2}^2
\label{eqnn9}
\end{equation}
\noindent
and
\begin{equation}
\label{masseq6}
m_{S2}^2 = m_S'^2 + {\lambda_{11}}c_1^2 + {\lambda_{12}}(c_2^2 + c_3^2) + \frac{1}{4}\lambda_{10} c_4^2.
\end{equation}

\noindent
Furthermore, in order to be stable against ${\cal CB}_p$ vacuum, one requires

\noindent
\begin{equation}
{\cal V}_{{\cal CB}_{p}} - {\cal V}_{{\cal N}_{sp}} > 0,
\label{sp3}
\end{equation}

\noindent
where
\begin{equation}
{\cal V}_{{\cal CB}_{p}} - {\cal V}_{{\cal N}_{sp}} = (\frac{m_{H^{\pm}}^2}{2v^2})_{{\cal N}_{sp}}[(v_2 c_1 - v_1 c_3)^2 + v_1^2 c_2^2] - v_s^2 m_{S1}^2
\label{eqnn10}
\end{equation}

\noindent
and
\begin{equation}
\label{masseq7}
m_{S1}^2 = m_S^2 + {\lambda_{7}}c_1^2 + {\lambda_{8}}(c_2^2 + c_3^2) + \frac{1}{4}\lambda_{10} c_5^2.
\end{equation}

\noindent
Finally, for stability against ${\cal CB}_{sp}$ vacuum, one requires
\noindent
\begin{equation}
{\cal V}_{{\cal CB}_{sp}} - {\cal V}_{{\cal N}_{sp}} > 0,
\end{equation}
\noindent
where
\begin{equation}
{\cal V}_{{\cal CB}_{sp}} - {\cal V}_{{\cal N}_{sp}} = (\frac{m_{H^{\pm}}^2}{2v^2})_{{\cal N}_{sp}}[(v_2 c_1 - v_1 c_3)^2 + v_1^2 c_2^2].
\label{eqnnn10}
\end{equation}

\noindent
From Eq.~(\ref{eqnnn10}), it is clear that the stability condition is trivially satisfied in this case. Therefore, one can say that ${\cal N}_{sp}$ is absolutely stable against ${\cal CB}_{sp}$-type charge-breaking vacuum. However, its stability against ${\cal CB}$, ${\cal CB}_s$ and ${\cal CB}_p$-type charge-breaking minima is subject to the conditions Eqs.~(\ref{sp1}),~(\ref{sp2}) and (\ref{sp3}).

Finally, the corresponding conditions for ${\cal CP}$-breaking breaking vacua can be obtained by applying the following transformation in Eqs.~(\ref{eqnn8}),~(\ref{masseq5}),~(\ref{eqnn9}),~(\ref{masseq6}),~(\ref{eqnn10}),~(\ref{masseq7}) and~(\ref{eqnnn10}).

\begin{equation}
    c_1 \rightarrow c_1', c_3 \rightarrow c_2', c_2 \rightarrow c_3', c_4 \rightarrow c_4', c_5 \rightarrow c_5'.
\end{equation}

\subsection{Coexisting neutral minima or panic vacuum}

Apart from the dangerous charge and CP-breaking minima discussed above, one can also encounter ``wrong" neutral minima. For example, both doublets obtaining vev $v_1'$ and $v_2'$ respectively, where $\sqrt{v_1'^2 + v_2'^2} \neq 246$ GeV, or a situation where both doublets acquire 0 vev and only the singlet acquires vev. Such type of vacua if lower than the EW vacuum, will again lead to unphysical solutions. Therefore, it is important to also examine the conditions of stability of our EW vacua against these potentially dangerous neutral vacua. We define them as follows.

\begin{equation}
{\cal N'} \rightarrow
\braket{\Phi_1}_0=\frac{1}{\sqrt{2}}\left(\begin{array}{c} 0 \\
\,v_1'
\end{array}\right)\,, \ \ \
\braket{\Phi_2}_0=\frac{1}{\sqrt{2}}\left(\begin{array}{c} 0 \\
\,v_2'
\end{array}\right)\,, \ \ \
\braket{S}_0=0, \ \ \ \braket{P}_0=0,
\end{equation}

\begin{equation}
{\cal N}_s' \rightarrow
\braket{\Phi_1}_0=\frac{1}{\sqrt{2}}\left(\begin{array}{c} 0 \\
\,v_1'
\end{array}\right)\,, \ \ \
\braket{\Phi_2}_0=\frac{1}{\sqrt{2}}\left(\begin{array}{c} 0 \\
\,v_2'
\end{array}\right)\,, \ \ \
\braket{S}_0=s', \ \ \ \braket{P}_0=0,
\end{equation}

\begin{equation}
{\cal N}_p' \rightarrow
\braket{\Phi_1}_0=\frac{1}{\sqrt{2}}\left(\begin{array}{c} 0 \\
\,v_1'
\end{array}\right)\,, \ \ \
\braket{\Phi_2}_0=\frac{1}{\sqrt{2}}\left(\begin{array}{c} 0 \\
\,v_2'
\end{array}\right)\,, \ \ \
\braket{S}_0=0, \ \ \ \braket{P}_0=p',
\end{equation}

\begin{equation}
{\cal N}_{sp}' \rightarrow
\braket{\Phi_1}_0=\frac{1}{\sqrt{2}}\left(\begin{array}{c} 0 \\
\,v_1'
\end{array}\right)\,, \ \ \
\braket{\Phi_2}_0=\frac{1}{\sqrt{2}}\left(\begin{array}{c} 0 \\
\,v_2'
\end{array}\right)\,, \ \ \
\braket{S}_0=s', \ \ \ \braket{P}_0=p',
\end{equation}

\begin{equation}
{\cal S} \rightarrow
\braket{\Phi_1}_0=\frac{1}{\sqrt{2}}\left(\begin{array}{c} 0 \\
\,0
\end{array}\right)\,, \ \ \
\braket{\Phi_2}_0=\frac{1}{\sqrt{2}}\left(\begin{array}{c} 0 \\
\,0
\end{array}\right)\,, \ \ \
\braket{S}_0=s', \ \ \ \braket{P}_0=0,
\end{equation}

\begin{equation}
{\cal P} \rightarrow
\braket{\Phi_1}_0=\frac{1}{\sqrt{2}}\left(\begin{array}{c} 0 \\
\,0
\end{array}\right)\,, \ \ \
\braket{\Phi_2}_0=\frac{1}{\sqrt{2}}\left(\begin{array}{c} 0 \\
\,0
\end{array}\right)\,, \ \ \
\braket{S}_0=0, \ \ \ \braket{P}_0=p',
\end{equation}

\begin{equation}
{\cal {SP}} \rightarrow
\braket{\Phi_1}_0=\frac{1}{\sqrt{2}}\left(\begin{array}{c} 0 \\
\,0
\end{array}\right)\,, \ \ \
\braket{\Phi_2}_0=\frac{1}{\sqrt{2}}\left(\begin{array}{c} 0 \\
\,0
\end{array}\right)\,, \ \ \
\braket{S}_0=s', \ \ \ \braket{P}_0=p'.
\end{equation}

\noindent
The stability conditions for ${\cal N}$-, ${\cal N}_s$-, ${\cal N}_p$- and ${\cal N}_{sp}$-type vacua against possible {\it panic} vacua are given as follows:

\begin{equation}
{\cal V}_{\cal N'} - {\cal V}_{\cal N} = \left[\left(\frac{m_{H^{\pm}}^2}{2v^2}\right)_{\cal N} - \left(\frac{m_{H^{\pm}}^2}{2v^2}\right)_{\cal N'}\right] (v_2' v_1 - v_1' v_2)^2 > 0,
\label{eq100}
\end{equation} 

\begin{equation}
{{\cal V}_{\cal N}}_s' - {\cal V}_{\cal N} = \left[\left(\frac{m_{H^{\pm}}^2}{2v^2}\right)_{\cal N} - \left(\frac{m_{H^{\pm}}^2}{2v^2}\right)_{{\cal N}'_s}\right] (v_2' v_1 - v_1' v_2)^2 + m_s^2 s'^2 > 0,
\label{eq101}
\end{equation}

\begin{equation}
{{\cal V}_{\cal N}}_p' - {\cal V}_{\cal N} = \left[\left(\frac{m_{H^{\pm}}^2}{2v^2}\right)_{\cal N} - \left(\frac{m_{H^{\pm}}^2}{2v^2}\right)_{{\cal N}'_p}\right] (v_2' v_1 - v_1' v_2)^2 + m_p^2 p'^2 > 0,
\label{eq102}
\end{equation}

\begin{equation}
{{\cal V}_{\cal N}}_{sp}' - {\cal V}_{\cal N} = \left[\left(\frac{m_{H^{\pm}}^2}{2v^2}\right)_{\cal N} - \left(\frac{m_{H^{\pm}}^2}{2v^2}\right)_{{\cal N}'_{sp}}\right] (v_2' v_1 - v_1' v_2)^2 + m_s^2 s'^2 + m_p^2 p'^2 > 0,
\label{eq103}
\end{equation} 

\begin{equation}
{\cal V}_{\cal N'} - {{\cal V}_{\cal N}}_s = \left[\left(\frac{m_{H^{\pm}}^2}{2v^2}\right)_{{\cal N}_s} - \left(\frac{m_{H^{\pm}}^2}{2v^2}\right)_{\cal N'}\right] (v_2' v_1 - v_1' v_2)^2 - v_s^2 m_{S1}^2 > 0,
\label{eq104}
\end{equation} 

\begin{equation}
{\cal V}_{{\cal N'}_s} - {{\cal V}_{\cal N}}_s = \left[\left(\frac{m_{H^{\pm}}^2}{2v^2}\right)_{{\cal N}_s} - \left(\frac{m_{H^{\pm}}^2}{2v^2}\right)_{{\cal N}'_s}\right] (v_2' v_1 - v_1' v_2)^2 > 0,
\label{eq105}
\end{equation}

\begin{equation}
{\cal V}_{{\cal N'}_p} - {{\cal V}_{\cal N}}_s = \left[\left(\frac{m_{H^{\pm}}^2}{2v^2}\right)_{{\cal N}_s} - \left(\frac{m_{H^{\pm}}^2}{2v^2}\right)_{{\cal N}'_p}\right] (v_2' v_1 - v_1' v_2)^2 - v_s^2 m_{S1}^2 + m_p^2 p'^2 > 0,
\label{eq106}
\end{equation}

\begin{equation}
{\cal V}_{{\cal N'}_{sp}} - {{\cal V}_{\cal N}}_s = \left[\left(\frac{m_{H^{\pm}}^2}{2v^2}\right)_{{\cal N}_s} - \left(\frac{m_{H^{\pm}}^2}{2v^2}\right)_{{\cal N}'_{sp}}\right] (v_2' v_1 - v_1' v_2)^2 + m_p^2 p'^2 > 0,
\label{eq107}
\end{equation}

\begin{equation}
{\cal V}_{\cal N'} - {{\cal V}_{\cal N}}_p = \left[\left(\frac{m_{H^{\pm}}^2}{2v^2}\right)_{{\cal N}_p} - \left(\frac{m_{H^{\pm}}^2}{2v^2}\right)_{\cal N'}\right] (v_2' v_1 - v_1' v_2)^2 - v_p^2 m_{S2}^2 > 0,
\label{eq108}
\end{equation} 

\begin{equation}
{\cal V}_{{\cal N'}_s} - {{\cal V}_{\cal N}}_p = \left[\left(\frac{m_{H^{\pm}}^2}{2v^2}\right)_{{\cal N}_p} - \left(\frac{m_{H^{\pm}}^2}{2v^2}\right)_{{\cal N}'_s}\right] (v_2' v_1 - v_1' v_2)^2 - v_p^2 m_{S2}^2 + m_s^2 s'^2 > 0,
\label{eq109}
\end{equation}

\begin{equation}
{\cal V}_{{\cal N'}_p} - {{\cal V}_{\cal N}}_p = \left[\left(\frac{m_{H^{\pm}}^2}{2v^2}\right)_{{\cal N}_p} - \left(\frac{m_{H^{\pm}}^2}{2v^2}\right)_{{\cal N}'_p}\right] (v_2' v_1 - v_1' v_2)^2 > 0,
\label{eq110}
\end{equation}

\begin{equation}
{\cal V}_{{\cal N'}_{sp}} - {{\cal V}_{\cal N}}_p = \left[\left(\frac{m_{H^{\pm}}^2}{2v^2}\right)_{{\cal N}_p} - \left(\frac{m_{H^{\pm}}^2}{2v^2}\right)_{{\cal N}'_{sp}}\right] (v_2' v_1 - v_1' v_2)^2 + m_s^2 s'^2 > 0,
\label{eq111}
\end{equation}

\begin{equation}
{\cal V}_{\cal N'} - {{\cal V}_{\cal N}}_{sp} = \left[\left(\frac{m_{H^{\pm}}^2}{2v^2}\right)_{{\cal N}_{sp}} - \left(\frac{m_{H^{\pm}}^2}{2v^2}\right)_{\cal N'}\right] (v_2' v_1 - v_1' v_2)^2 - v_s^2 m_{S1}^2 - v_p^2 m_{S2}^2 > 0,
\label{eq108}
\end{equation} 

\begin{equation}
{\cal V}_{{\cal N'}_s} - {{\cal V}_{\cal N}}_{sp} = \left[\left(\frac{m_{H^{\pm}}^2}{2v^2}\right)_{{\cal N}_{sp}} - \left(\frac{m_{H^{\pm}}^2}{2v^2}\right)_{{\cal N}'_s}\right] (v_2' v_1 - v_1' v_2)^2 - v_p^2 m_{S2}^2 > 0
\label{eq109},
\end{equation}

\begin{equation}
{\cal V}_{{\cal N'}_p} - {{\cal V}_{\cal N}}_{sp} = \left[\left(\frac{m_{H^{\pm}}^2}{2v^2}\right)_{{\cal N}_{sp}} - \left(\frac{m_{H^{\pm}}^2}{2v^2}\right)_{{\cal N}'_p}\right] (v_2' v_1 - v_1' v_2)^2 - v_s^2 m_{S1}^2 > 0
\label{eq110},
\end{equation}

\begin{equation}
{\cal V}_{{\cal N'}_{sp}} - {{\cal V}_{\cal N}}_{sp} = \left[\left(\frac{m_{H^{\pm}}^2}{2v^2}\right)_{{\cal N}_{sp}} - \left(\frac{m_{H^{\pm}}^2}{2v^2}\right)_{{\cal N}'_{sp}}\right] (v_2' v_1 - v_1' v_2)^2 > 0.
\label{eq111}
\end{equation}

\noindent
One can see that, similar to the charge and CP-breaking scenarios, here too, some of the aforementioned conditions are trivially satisfied and the rest are subject to the values of the free parameters of the corresponding model. 

In the next section, we will explore the difference between N2HDM and 2HDMS numerically for a chosen benchmark. We mention here, that we choose a benchmark in the N2HDM of ${\cal N}_s$-type vacuum, and compare with a 2HDMS scenario with ${\cal N}_{sp}$ as well as ${\cal N}_s$-type vacuum. We do not compare ${\cal N}_{p}$-type vacuum of the 2HDMS separately, because in this case, the role of $S$ and $P$ are interchangeable and we expect similar results for ${\cal N}_s$ and ${\cal N}_p$-type vacuum of the 2HDMS. Furthermore, we do not make a comparison between ${\cal N}$-type vacua in both N2HDM and 2HDMS. The reasons are as follows. Firstly, it has been shown that in both models, the ${\cal N}$-type EW-vacuum is stable against charge or CP-breaking vacua. Secondly, one of our motivations behind studying the two models N2HDM and 2HDMS is the explanation of the recent 95 GeV excess available in both models. Such a state cannot be accommodated in the two models, with only ${\cal N}$-type vacuum.

\section{Exploring the difference for specific benchmarks: ${\cal N}_{sp}$-type vacuum}
\label{sec4}

Having discussed the different conditions for (in)stabilities in the 2HDMS and its comparison with those in the N2HDM, we explore this difference numerically first in terms of a couple of chosen benchmark points. We will first consider the case of ${\cal N}_{sp}$-type vacuum, i.e both components of the complex singlet field namely, $S$ and $P$ acquire a vev in this case. For the numerical study we use the code {\tt EVADE}~\cite{Hollik:2018wrr,Ferreira:2019iqb}. {\tt EVADE} finds
the tree-level minima via employing {\tt HOM4PS2}. Furthermore, in case of the EW vacuum being a false vacuum, it
calculates the bounce action with a straight path approximation.

\subsection{Fate of a stable point}
\label{stable_n2hdm}

We first choose a stable benchmark of the N2HDM, presented in Tables~\ref{bp1_massbasis} and \ref{bp1_genbasis}, and explore the scenario by varying the additional free parameters of the 2HDMS freely.

{\scriptsize{
\begin{table}[!hptb]
\begin{center}
\begin{tabular}{|c|c|c|c|c|c|c|c|c|c|c|}
\hline
 & $m_{h_1}$ & $m_{h_2}$ & $m_{h_3}$ & $m_A=m_H^{\pm}$ & $m_{12}^2$ & $\tan\beta$ & $v_s$ & $\{\alpha_1,\alpha_2,\alpha_3\}$ \\
\hline
BP1 & 95.4 & 125 & 601 & 621 & 9529.17 & 1.37  &  468.1  &  \{-0.49,0.31,-0.09\} \\
\hline
\end{tabular}
\caption{Stable benchmark scenario of  the N2HDM, BP1 in the mass-basis.}
\label{bp1_massbasis}
\end{center}
\end{table}
}}

{\scriptsize{
\begin{table}[!hptb]
\begin{center}
\begin{tabular}{|c|c|c|c|c|c|c|c|c|c|c|}
\hline
 & $\lambda_1$ & $\lambda_2$ & $\lambda_3$ & $\lambda_4=\lambda_5$ & $\lambda_6$ & $\lambda_7$ & $\lambda_8$ & $m_{12}^2$ & $\tan\beta$ & $v_s$ \\
\hline
BP1 & 1.43 & 0.24 & 12.02 & -6.05 & 2.97 & 2.11 & -0.41 & 9529.17 & 1.37  &  468.1 \\
\hline
\end{tabular}
\caption{Stable benchmark scenario of the N2HDM BP1 in the interaction-basis.}
\label{bp1_genbasis}
\end{center}
\end{table}
}}

\noindent
We can see from Tables~\ref{bp1_massbasis} and \ref{bp1_genbasis} that BP1 accommodates the recently observed 95 GeV di-photon excess. The ATLAS experiments saw an excess of events around a mass 95.4 GeV with 1.7$\sigma$~\cite{ATLAS-CONF-2018-025} signal significance with a signal strength given as follows:

\begin{equation}
    \mu^{\text ATLAS}_{\gamma\gamma} = \frac{\sigma^{\text exp}(pp \rightarrow \phi \rightarrow \gamma\gamma)}{\sigma^{\text exp}(pp \rightarrow H \rightarrow \gamma\gamma)} = 0.18^{+0.10}_{-0.10}.
\end{equation}

\noindent
CMS, on the other hand, reported the observed access around 95.4 GeV with a signal significance of 2.9$\sigma$~\cite{CMS-PAS-HIG-20-002} with a given signal strength of

\begin{equation}
    \mu^{\text CMS}_{\gamma\gamma} = \frac{\sigma^{\text exp}(pp \rightarrow \phi \rightarrow \gamma\gamma)}{\sigma^{\text exp}(pp \rightarrow H \rightarrow \gamma\gamma)} = 0.33^{+0.19}_{-0.12}.
\end{equation}

\noindent
In our study here, we use the combined result~\cite{Biekotter:2023oen} of these excesses assuming no correlation. The combined signal strength is obtained as.

\begin{equation}
    \mu^{\text exp}_{\gamma\gamma} = \mu^{\text CMS+ATLAS}_{\gamma\gamma} = 0.24^{+0.09}_{-0.08},
\end{equation}

\noindent
which corresponds to an excess of 3.1 $\sigma$ at 95.4 GeV.
Alongside, the benchmark also satisfies all the experimental observations of the 125 GeV SM-like Higgs as well as the bounds for the non-standard  scalars. The aforementioned constraints are checked with {\tt HiggsTools}~\cite{Bahl:2022igd,Bechtle:2013xfa,Bechtle:2013wla,Bechtle:2020pkv,Bechtle:2020uwn}.

\begin{figure}[!hptb]
	\centering
	\includegraphics[width=6.0cm,height=5cm] 
        {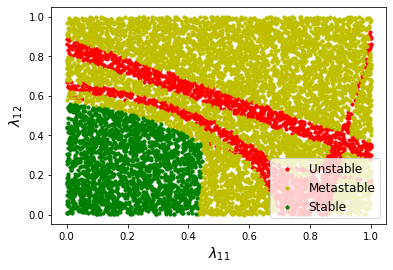}
        \includegraphics[width=6.0cm,height=5cm]{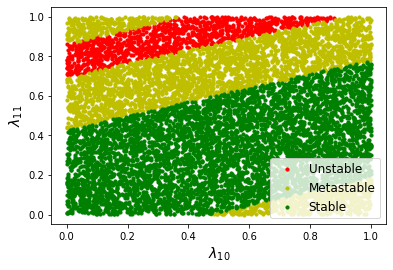} \\
        \includegraphics[width=6.0cm,height=5cm]{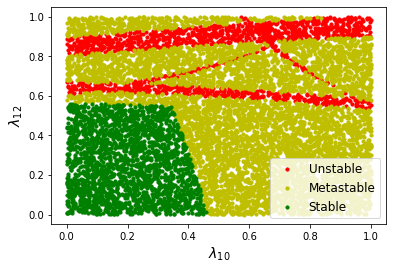}
	\caption{BP1 : The stable(dark green), metastable(light green) and unstable(red) points in the 2HDMS}
	\label{stable_bp}
\end{figure}

We have also checked that the N2HDM benchmark satisfies the bounded-from-below conditions~\cite{Dutta:2023cig}. Furthermore, the stability of the BP has been checked with {\tt EVADE}, and this chosen benchmark turns out to be absolutely stable. Next we explore the stability of this benchmark point in the 2HDMS, i.e, varying the extra free parameters of the model, namely the quartic couplings ($\lambda_9,\lambda_{10},\lambda_{11}, \lambda_{12}$) within the perturbative limit, while the rest of the parameters are kept fixed at their corresponding values in the N2HDM in the interaction basis, as given in  Table~\ref{bp1_genbasis}. We will now present the parameter spaces for these four additional quartic couplings in the 2HDMS for this specific benchmark in Figure~\ref{stable_bp}. In all these figures at one point only two of the 2HDMS quartic couplings have been assumed to be non-zero and the other two are kept at zero. Therefore, the origin (0,0) in all these plots correspond to the N2HDM limit of the 2HDMS parameter space.

We can clearly see from Figure~\ref{stable_bp} that when the quartic couplings are 0 (in the N2HDM limit) or small, the stability is intact. However, when the additional parameters are allowed to vary freely in 2HDMS, large values of them can make a stable point of the N2HDM metastable and even unstable. This behavior can be understood by using the analytical expressions for the stability conditions for the parameter point (see Section~\ref{sec3}). The kinks in the distributions can be understood as the points where different conditions start having impact on the stability of the parameter space.

\subsection{Fate of a metastable point}
\label{metastable_n2hdm}

Next, we choose a metastable benchmark in the N2HDM and repeat the same procedure as in Subsection~\ref{stable_n2hdm}. The chosen BP satisfies all the aforementioned (in Subsection~\ref{stable_n2hdm}) theoretical and experimental constraints and accommodates the combined result of ATLAS and CMS for the excess at 95.4 GeV.

{\scriptsize{
\begin{table}[!hptb]
\begin{center}
\begin{tabular}{|c|c|c|c|c|c|c|c|c|c|c|}
\hline
 & $m_{h_1}$ & $m_{h_2}$ & $m_{h_3}$ & $m_A=m_H^{\pm}$ & $m_{12}^2$ & $\tan\beta$ & $v_s$ & $\{\alpha_1,\alpha_2,\alpha_3\}$ \\
\hline
BP2 & 95.4 & 125 & 607.8 & 628.0 & -13222.9 & 1.48  &  286.1  &  \{-0.45,0.86,-0.09\} \\
\hline
\end{tabular}
\caption{Metastable benchmark scenario BP2 of the N2HDM in the mass-basis.}
\label{bp2_massbasis}
\end{center}
\end{table}
}}

{\scriptsize{
\begin{table}[!hptb]
\begin{center}
\begin{tabular}{|c|c|c|c|c|c|c|c|c|c|c|}
\hline
 & $\lambda_1$ & $\lambda_2$ & $\lambda_3$ & $\lambda_4=\lambda_5$ & $\lambda_6$ & $\lambda_7$ & $\lambda_8$ & $m_{12}^2$ & $\tan\beta$ & $v_s$ \\
\hline
BP2 & 12.44 & 0.58 & 12.84 & -6.99 & 3.99 & 6.35  & -0.30 & -13222.9 & 1.48 & 286.1 \\
\hline
\end{tabular}
\caption{Metastable benchmark scenario BP2 of the N2HDM in the interaction-basis.}
\end{center}
\label{bp2_genbasis}
\end{table}
}}

\begin{figure}[!hptb]
	\centering
	\includegraphics[width=6.0cm,height=5cm]{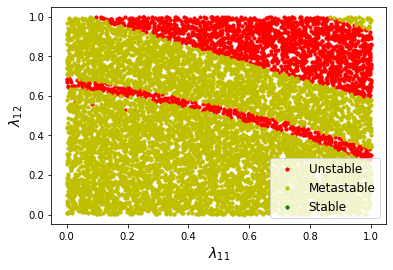}
   	\includegraphics[width=6.0cm,height=5cm]{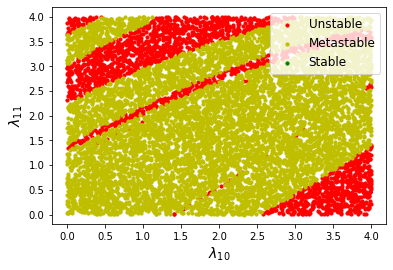} \\
      \includegraphics[width=6.0cm,height=5cm]{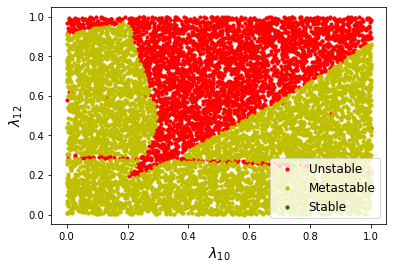} 
	\caption{BP2 : The stable(dark green), metastable(light green) and unstable(red) points in the 2HDMS.}
	\label{metastable_bp}
\end{figure}

\noindent
In Figure~\ref{metastable_bp}, similar as in the previous case, only two quartic couplings are varied simultaneously and the others have been kept fixed at 0. Here too, the origin (0,0) of the plots corresponds to the N2HDM limit of the 2HDMS and is therefore metastable. One can see that with larger values of $\lambda_{11}$ and $\lambda{12}$, one can encounter unstable regions. Please note that the metastability and instability conditions are evaluated numerically using {\tt EVADE} and EVADE uses straight path approximation while calculating the transition probability of the EW minima to a potentially dangerous minima.

\subsection{Fate of an unstable point}

In our next example we choose an unstable benchmark of N2HDM, satisfying all theoretical and experimental constraints as well as accommodating the combined ATLAS+CMS excess at 95.4 GeV. We explore its vacuum (in)stability in the 2HDMS.

{\scriptsize{
\begin{table}[!hptb]
\begin{center}
\begin{tabular}{|c|c|c|c|c|c|c|c|c|c|c|}
\hline
 & $m_{h_1}$ & $m_{h_2}$ & $m_{h_3}$ & $m_A=m_H^{\pm}$ & $m_{12}^2$ & $\tan\beta$ & $v_s$ & $\{\alpha_1,\alpha_2,\alpha_3\}$ \\
\hline
BP3 & 95 & 125 & 614.3 & 634.3 & 17856.8 & 1.82  &  441.42  &  \{-0.50,0.37,0.05\} \\
\hline
\end{tabular}
\caption{Unstable benchmark scenario BP3 of N2HDM in the mass-basis}
\end{center}
\label{bp5}
\end{table}
}}

{\scriptsize{
\begin{table}[!hptb]
\begin{center}
\begin{tabular}{|c|c|c|c|c|c|c|c|c|c|c|}
\hline
 & $\lambda_1$ & $\lambda_2$ & $\lambda_3$ & $\lambda_4=\lambda_5$ & $\lambda_6$ & $\lambda_7$ & $\lambda_8$ & $m_{12}^2$ & $\tan\beta$ & $v_s$ \\
\hline
BP3 & 1.93 & 0.11 & 12.72  & -5.95 & 3.36 & 3.38 & 0.23 & 17856.8 & 1.82 & 441.42 \\
\hline
\end{tabular}
\caption{Unstable benchmark scenario BP3 of N2HDM in the interaction-basis}
\end{center}
\label{bp5}
\end{table}
}}

\begin{figure}[!hptb]
	\centering
	\includegraphics[width=6.0cm,height=5cm]{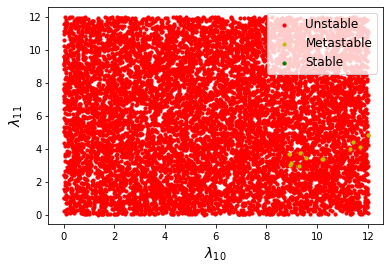}
        \includegraphics[width=6.0cm,height=5cm]{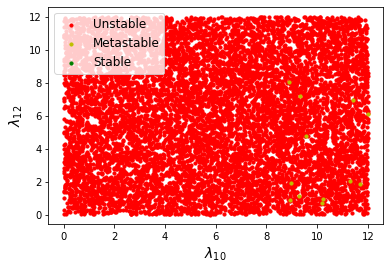} \\
        \includegraphics[width=6.0cm,height=5cm]{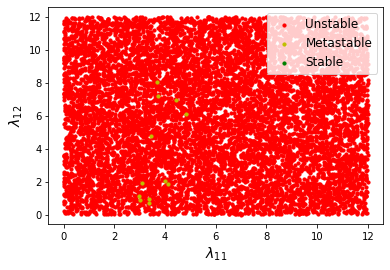} 
        \caption{BP3 : The stable(dark green), metastable(light green) and unstable(red) points in the 2HDMS.}
	\label{unstable_bp}
\end{figure}

It is clearly seen from Figure~\ref{unstable_bp}, that the unstable point of N2HDM remains mostly unstable. It can become metastable only for extremely fine-tuned regions of parameters denoted in light green points in Figure~\ref{unstable_bp}. 

\noindent
Pertaining to the discussion so far, we can summarize. We saw that, a stable point of N2HDM can become metastable or unstable in 2HDMS. But the converse is disfavored. The reason behind that is a follows. 

\begin{itemize}
\item We have already discussed that the number of possible dangerous vacua is higher in case of the 2HDMS compared to the N2HDM. For example, if we compare ${\cal N}_s$-type vacua in both cases, we see that, ${\cal N}_s$-type vacua can be unstable against ${\cal CB}$-type vacuum in the N2HDM(subject to condition Eq.~(\ref{eqn1})). But in the 2HDMS the vacuum can be unstable against ${\cal CB}$ as well as ${\cal CB}_{p}$-type vacuum(subject to Eqs. (\ref{eqn1}) and (\ref{eqnn11}), respectively). On the other hand, if we compare ${\cal N}_{sp}$-type vacua of the 2HDMS with ${\cal N}_s$-type vacua of the N2HDM, we see that ${\cal N}_{sp}$-type vacuum of the 2HDMS can be unstable against ${\cal CB}$-, ${\cal CB}_{s}$- and ${\cal CB}_{p}$-type vacuum (subject to conditions in Eqs. (\ref{sp1}), (\ref{sp2}) and (\ref{sp3}) respectively), while ${\cal N}_s$-type vacuum of the N2HDM is unstable against only ${\cal CB}$-type vacuum (subject to condition in Eq.~(\ref{eqn1})). So it is evident that if the single stability condition is satisfied in the N2HDM for a specific benchmark point, it is possible that if any of the multiple stability conditions in the 2HDMS is not satisfied and the point becomes metastable/unstable. It can, in particular, happen for large positive values of couplings $\lambda_{10}$, $\lambda_{11}$ and $\lambda_{12}$ (see the corresponding Equations).  
\item On the other hand, concerning a metastable or unstable point of the N2HDM, it has to satisfy all the stability criteria, i.e. Eqs. (\ref{eqn1}) and (\ref{eqnn11}) for ${\cal N}_s$-type vacuum and all three conditions (Eqs. (\ref{sp1}), (\ref{sp2}) and (\ref{sp3})) for ${\cal N}_{sp}$-vacuum, in order to become stable in the 2HDMS framework, which is evidently difficult. In case of ${\cal N}_s$-type vacuum it is absolutely impossible, since Eq. (\ref{eqn1}) is already violated in case of an unstable point in the N2HDM. In case of ${\cal N}_{sp}$ vacuum, it is in principle possible to fulfill simultaneously the mentioned stability conditions, but it will require large negative values of the couplings $\lambda_{10}, \lambda_{11}$ and $\lambda_{12}$, i.e, a scenario which is highly constrained from the requirement that the potential should be bounded from below. In order to summarize this part we present Table~\ref{n2hdm_2hdms_vacuua}, where the possible fate of N2HDM vacua under 2HDMS extensions of the scalar potential is listed.

\begin{table}
\scriptsize{\begin{tabular}{| c |c | c |c |c | }
\hline
\multicolumn{1}{|c|}{} &
\multicolumn{2}{|c|}{${\cal N}_s$ vacuum} &
\multicolumn{2}{|c|}{${\cal N}_{sp}$ vacuum}\\
\hline
\multicolumn{1}{|c|}{N2HDM} &
\multicolumn{1}{|c|}{stable} &
\multicolumn{1}{|c|}{meta/unstable} &
\multicolumn{1}{|c|}{stable} &
\multicolumn{1}{|c|}{meta/unstable} \\
\multicolumn{1}{|c|}{} &
\multicolumn{1}{|c|}{$\downarrow$} &
\multicolumn{1}{|c|}{$\downarrow$} &
\multicolumn{1}{|c|}{$\downarrow$} &
\multicolumn{1}{|c|}{$\downarrow$} \\ 
\hline
\multicolumn{1}{|c|}{2HDMS} &
\multicolumn{1}{|c|}{stable/meta/unstable} &
\multicolumn{1}{|c|}{meta/unstable} &
\multicolumn{1}{|c|}{stable/meta/unstable} &
\multicolumn{1}{|c|}{stable(fine-tuned)/meta/unstable} \\
\hline
\end{tabular}}
\caption{The stable and meta/unstable vacuum in N2HDM can lead to stable or meta/unstable vacuum in 2HDMS depending on the type of vacuum ${\cal N}_s$ or ${\cal N}_{sp}$.}
\label{n2hdm_2hdms_vacuua}
\end{table}

\end{itemize}

\section{Comparison between N2HDM and 2HDMS with observational constraints: ${\cal N}_{sp}$ vacuum }
\label{sec5}

In the discussion so far, we have seen that the vacuum stability of a parameter point in the N2HDM, can be quite different compared to in the 2HDMS, depending on the extra parameters of the model. However, as soon as some experimental evidences for BSM physics have been made and the parameter regions in the different models will be strongly constrained leaving also less variability for the vacuum structure in both models.
Therefore, we want to further explore the differences in the vacuum structures of both models, but based on possibly made BSM measurements of masses and couplings.

\subsection{Impact of Yukawa and gauge boson couplings of scalars}

In order to make a comparison between the two models in terms of physical observables, we take up the following strategy. We first assume that new BSM Higgs particles have been found and try to match the mass pattern in both models and thereby construct the relevant mixing matrices. 
Since all the couplings of physical scalars are functions of scalar mixing matrix elements, we demand all the elements of the 3$\times$3 subspace of the 4$\times$4 mixing matrix of 2HDMS to be within $\lsim$ 15\% of the matrix elements of the 3$\times$3 mixing matrix of N2HDM. In other words, $\Delta C/C \lsim 15\%$ where $C$ is a generic Yukawa as well as gauge boson couplings of $h_1$, $h_2$ or $h_3$ and $\Delta C$ is the difference between their respective values in the N2HDM and 2HDMS.

We write down the mixing matrices of the N2HDM and the 2HDMS in terms of three and six mixing angles respectively. We denote the three mixing angles of the N2HDM as $\alpha_{1,..3}$ and the six mixing angles of the 2HDMS as $\alpha_{1,..6}$. 

The 3$\times 3$ mixing matrix R in the basis as in Eq.~(\ref{n2hdm_basis}) of the N2HDM is given as follows. 

\begin{equation}
 R = 
 \begin{pmatrix}
 \scriptsize
  R_{11}  & R_{12} & R_{13} \\
  R_{21} & R_{22} & R_{23} \\
  R_{31} & R_{32} & R_{33}\\
 \end{pmatrix},
\end{equation}

where the $R_{ij}$'s are parametrized as

\begin{eqnarray}
 R_{11} &=& c_{\alpha_1}  c_{\alpha_2}, \nonumber \\
 R_{12} &=& s_{\alpha_1} c_{\alpha_2}, \nonumber \\
 R_{13} &=& s_{\alpha_2}, \nonumber \\
 R_{21} &=& -s_{\alpha_1} c_{\alpha_3} - c_{\alpha_1} s_{\alpha_2} s_{\alpha_3},  \nonumber \\
 R_{22} &=& c_{\alpha_1} c_{\alpha_3} - s_{\alpha_1} s_{\alpha_2} s_{\alpha_3}, \nonumber \\  
 R_{23} &=& c_{\alpha_2} s_{\alpha_3} \nonumber \\
 R_{31} &=& s_{\alpha_1} s_{\alpha_3} - c_{\alpha_1} s_{\alpha_2} c_{\alpha_3}, \nonumber \\
 R_{32} &=& - c_{\alpha_1} s_{\alpha_3 }- s_{\alpha_1} s_{\alpha_2} c_{\alpha_3}, \nonumber \\
 R_{33} &=& c_{\alpha_2} c_{\alpha_3}.
\end{eqnarray}

The corresponding mixing matrix elements in case of the 4$\times$4 mixing matrix $R'$ in 2HDMS as in Eq.~(\ref{2hdms_basis}) are as follows.

\begin{equation}
 R' = 
 \begin{pmatrix}
 \scriptsize
  R'_{11}  & R'_{12} & R'_{13} & R'_{14} \\
  R'_{21} & R'_{22} & R'_{23} & R'_{24} \\
  R'_{31} & R'_{32} & R'_{33} & R'_{34} \\
  R'_{41} & R'_{42} & R'_{43} & R'_{44}
 \end{pmatrix},
\end{equation}

where the components $R'_{ij}$'s are given as

\begin{eqnarray*}
R'_{11} &=& c_{\alpha_1}  c_{\alpha_2} c_{\alpha_4}, \\
R'_{12} &=& c_{\alpha_2} c_{\alpha_4} s_{\alpha_1},\\
R'_{13} &=& c_{\alpha_4} s_{\alpha_2},\\
R'_{14} &=& s_{\alpha_4},\\
R'_{21} &=& c_{\alpha_5}(-c_{\alpha_3} s_{\alpha_1} - c_{\alpha_1} s_{\alpha_2} s_{\alpha_3}) - c_{\alpha_1} c_{\alpha_2} s_{\alpha_4} s_{\alpha_5}, \\
R'_{22} &=& c_{\alpha_5}(c_{\alpha_1} c_{\alpha_3} - s_{\alpha_1} s_{\alpha_2} s_{\alpha_3}) - c_{\alpha_2} s_{\alpha_1} s_{\alpha_4} s_{\alpha_5},\\
R'_{23} &=& c_{\alpha_2} c_{\alpha_5} s_{\alpha_3} - s_{\alpha_2} s_{\alpha_4} s_{\alpha_5},  \\
R'_{24} &=& c_{\alpha_4} s_{\alpha_5},\\
R'_{31} &=& c_{\alpha_6} (-c_{\alpha_1} c_{\alpha_3} s_{\alpha_2} + s_{\alpha_1} s_{\alpha_3}) + (-c_{\alpha_1} c_{\alpha_2} c_{\alpha_5} s_{\alpha_4} - (-c_{\alpha_3} s_{\alpha_1} - c_{\alpha_1} s_{\alpha_2} s_{\alpha_3}) s_{\alpha_5}) s_{\alpha_6},\\
R'_{32} &=& c_{\alpha_6} (-s_{\alpha_1} c_{\alpha_3} s_{\alpha_2} - c_{\alpha_1} s_{\alpha_3}) + (-s_{\alpha_1} c_{\alpha_2} c_{\alpha_5} s_{\alpha_4} - (-c_{\alpha_3} c_{\alpha_1} - s_{\alpha_1} s_{\alpha_2} s_{\alpha_3}) s_{\alpha_5}) s_{\alpha_6},\\
R'_{33} &=& c_{\alpha_2} c_{\alpha_3} c_{\alpha_6} + (-c_{\alpha_5} s_{\alpha_2} s_{\alpha_4} - c_{\alpha_2} s_{\alpha_3} s_{\alpha_5})s_{\alpha_6}, \\
R'_{34} &=& c_{\alpha_4} c_{\alpha_5} s_{\alpha_6},\\
R'_{41} &=& c_{\alpha_6} (-c_{\alpha_1} c_{\alpha_2} c_{\alpha_5} s_{\alpha_4} - (-c_{\alpha_3} s_{\alpha_1} - c_{\alpha_1} s_{\alpha_2} s_{\alpha_3}) s_{\alpha_5}) -(-c_{\alpha_1} s_{\alpha_2} c_{\alpha_3} + s_{\alpha_1} s_{\alpha_3}) s_{\alpha_6},  \\
R'_{42} &=& c_{\alpha_6} (-s_{\alpha_1} c_{\alpha_2} c_{\alpha_5} s_{\alpha_4} - (c_{\alpha_3} c_{\alpha_1} - s_{\alpha_1} s_{\alpha_2} s_{\alpha_3}) s_{\alpha_5}) -(-s_{\alpha_1} s_{\alpha_2} c_{\alpha_3} - c_{\alpha_1} s_{\alpha_3}) s_{\alpha_6}, \\
R'_{43} &=& c_{\alpha_6} (-c_{\alpha_5} s_{\alpha_2} s_{\alpha_4} - c_{\alpha_2} s_{\alpha_3} s_{\alpha_5}) - c_{\alpha_2} c_{\alpha_3} s_{\alpha_6} \\
R'_{44} &=& c_{\alpha_4} c_{\alpha_5} c_{\alpha_6}. \\
\end{eqnarray*}

We further assume that all masses in both models are also close to each other within 10\% variation. We would also like to stress that we are not only imposing the existing experimental constraints such as the 125 GeV Higgs mass, its signal strengths etc. but we are furthermore assuming that all the extra scalars, which are not yet observed, but exist in both models and likely to be discovered in future experiments. Here, we demand the masses and signal strengths of all the scalars to be in the similar ballpark of each other. As discussed before, this in turn means that all the masses and the mixing angles in both models are close to each other. In this subsection, we will focus only on the Yukawa couplings and the gauge boson coupling of the scalars.

We emphasize that, this is, concerning the observable signal strengths, a rather conservative approach and we have only very limited freedom in choosing the extra mixing angles of the 2HDMS. Consequently, we are confined with small changes in the mixing angles.  
We will demonstrate that, even with such conservative approach, we see a differences between the N2HDM and the 2HDMS in terms of vacuum stability based on same physical observables($\lsim$ 15\% deviation being allowed).

For our analysis, we chose BP1 (presented in Table~\ref{bp1_massbasis}). We fix the masses and mixing angles as in Table~\ref{bp1_massbasis} for the N2HDM, and demand for the 2HDMS that the masses and mixing angles to vary within 15\% of those N2HDM values. We let $m_{12}^2$, vev's of the singlet scalars to be varied freely. One should note that, since we are working in the mass basis, we cannot vary the quartic couplings as input parameters. However, the variation of parameters in the mass basis, will in turn correspond to different regions in the parameter space of quartic couplings for the two models.

\begin{figure}
\centering
    \begin{tabular}{c c}
        \textbf{N2HDM} & \textbf{N2HDM} \\
        \includegraphics[width=6.7cm,height=5.5cm]{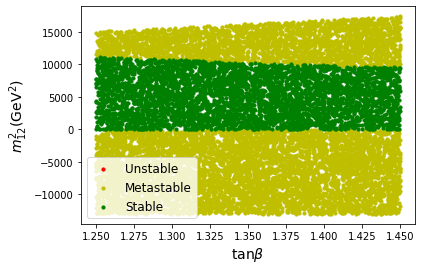}  &  \includegraphics[width=6.7cm,height=5.5cm]{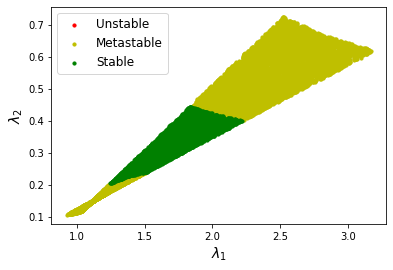}
    \end{tabular}
\caption{Allowed parameter space in $m_{12}^2-\tan\beta$ (left panel) and $\lambda_1-\lambda_2$ (right panel) plane in the N2HDM, when the scalar masses and mixing angles are kept fixed as in BP1 in Table~\ref{bp1_massbasis}.}
\label{bp1_0p2_scan_n2hdm}
\end{figure}

\begin{figure}
\centering
    \begin{tabular}{c c}
        \textbf{2HDMS} & \textbf{2HDMS} \\
        \includegraphics[width=6.7cm,height=5.5cm]{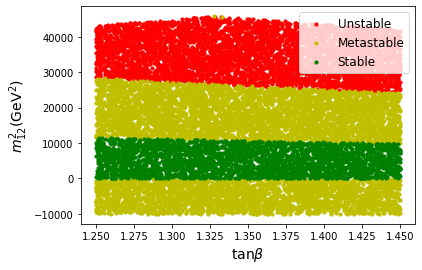}  &  \includegraphics[width=6.7cm,height=5.5cm]{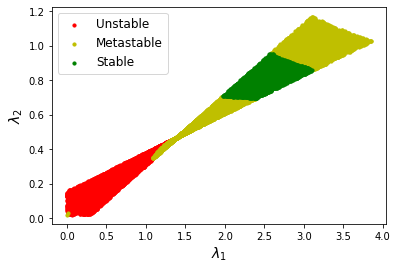}
    \end{tabular}
\caption{Allowed parameter space in $m_{12}^2-\tan\beta$ (left panel) and $\lambda_1-\lambda_2$ (right panel) plane in the 2HDMS, here the $h_1, h_2, h_3$ masses and mixing angles ($\alpha_1,\alpha_2,\alpha_3$) are kept fixed as in BP1. $\Delta C/C \lesssim 15\%$ which implies  $\alpha_4 \approx \alpha_5 \approx \alpha_6 \lesssim 0.2$. We assume $m_{h_4} = 700$ GeV and $v_p = 300$GeV.}
\label{bp1_0p2_scan_2hdms}
\end{figure}

In Figure,~\ref{bp1_0p2_scan_n2hdm}(left panel), we show the 
allowed region in $m_{12}^2-\tan\beta$ plane in the N2HDM. In Figure.~\ref{bp1_0p2_scan_n2hdm} (right panel), we show the allowed region in 
$\lambda_1-\lambda_2$ plane. The variation in the $\lambda_1-\lambda_2$ 
plane is not a result of an independent scanning of those parameters but a 
result of the variation of $m_{12}^2$, while keeping the physical masses
fixed. In Figure~\ref{bp1_0p2_scan_2hdms}, we present the analogous regions
in the 2HDMS. In order to adhere to $\lsim 15\%$ variation from the N2HDM, the 
additional mixing angles of the 2HDMS, namely $\alpha_4, \alpha_5$ and $\alpha_6$ are 
kept $\lsim 0.2$. By comparison, we can see that, we encounter 
unstable regions in the 2HDMS, while in case of the N2HDM, the allowed points are 
either stable or metastable. One should note that the parameter $m_{12}^2$
is not constrained by the observed signal strengths involving Yukawa or 
gauge couplings. It receives an upper limit from the boundedness from below 
constraint and the lower bound from perturbative unitarity, since all the 
physical masses are fixed. We see from Figures~\ref{bp1_0p2_scan_n2hdm} and
\ref{bp1_0p2_scan_2hdms}, that larger values of $m_{12}^2$ and consequently
smaller values of $\lambda_{1,2}$, are allowed in the 2HDMS, from the 
theoretical considerations, even with small difference in observed masses and
signal strengths. This leads to the possibility of unstable regions in the
2HDMS, unlike the N2HDM, when the masses and observed signal strengths are 
similar in both models.

\subsection{Impact of trilinear coupling measurement}

\begin{table}[!hptb]
\label{trilinear}
\begin{center}
\begin{tabular}{|c|c|c|c|c|}
\hline
Experiment & $\int{\cal L}dt$ & $\sqrt{s}$  & ${\cal P}[\%](e^-/e^+)$ & Expected Precision \\
\hline
\hline
HL-LHC & 14 TeV & 6 $ab^{-1}$ & -- & 50\%\\
\hline
HE-LHC & 27 TeV & 15 $ab^{-1}$ & -- & 10-20\%\\
\hline
FCC-hh & 100 TeV & 30 $ab^{-1}$ & -- & 5\% \\
\hline
ILC & 500 GeV  & 4 $ab^{-1}$ & $\pm80/\pm30$ & 27\% \\
\hline
ILC  & 1 TeV & 8 $ab^{-1}$ & $\pm80/\pm30$ & 10\% \\
\hline
CLIC & 3 TeV & 5 $ab^{-1}$  & $\pm80/0$ & 11\% \\
\hline
Muon collider & 14 TeV & 20 $ab^{-1}$ & -- & 3\%   \\
\hline
\end{tabular}
\caption{Expected precision of $\kappa_{\lambda}$ measurements at various future collider experiments~\cite{deBlas:2019rxi}.}
\end{center}
\label{trilinear}
\end{table}

\begin{figure}
\centering
    \begin{tabular}{c}
        \textbf{N2HDM} \\
        \includegraphics[width=7.5cm,height=5cm]{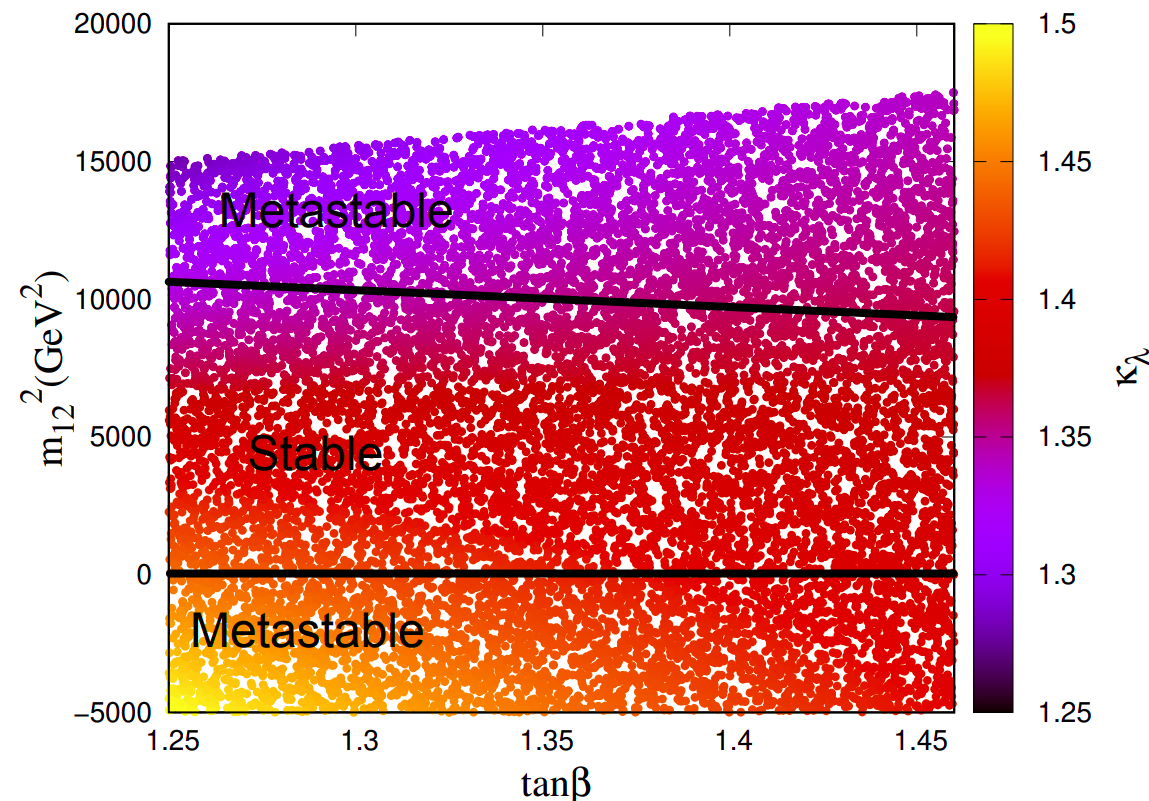}
        \end{tabular}
    \begin{tabular}{c c}
        \textbf{2HDMS} & \textbf{2HDMS} \\
        \includegraphics[width=7.5cm,height=5cm]  {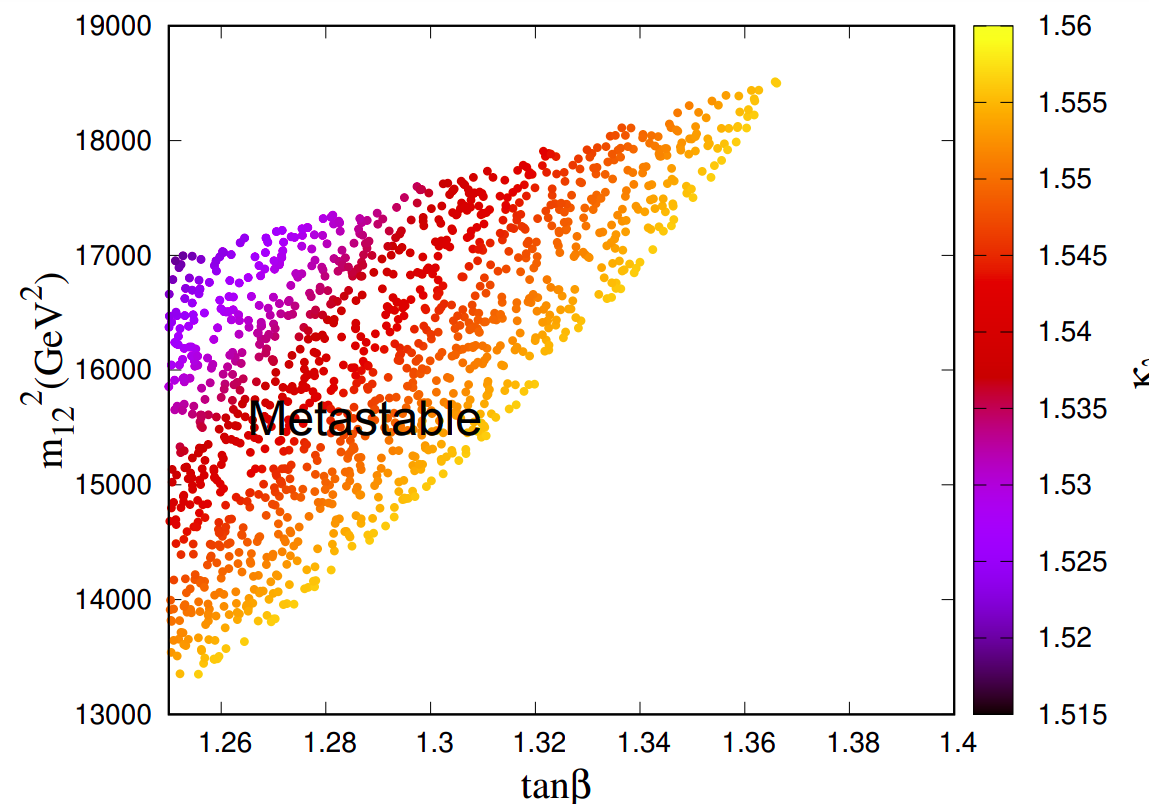} &
        \includegraphics[width=7.5cm,height=5cm]{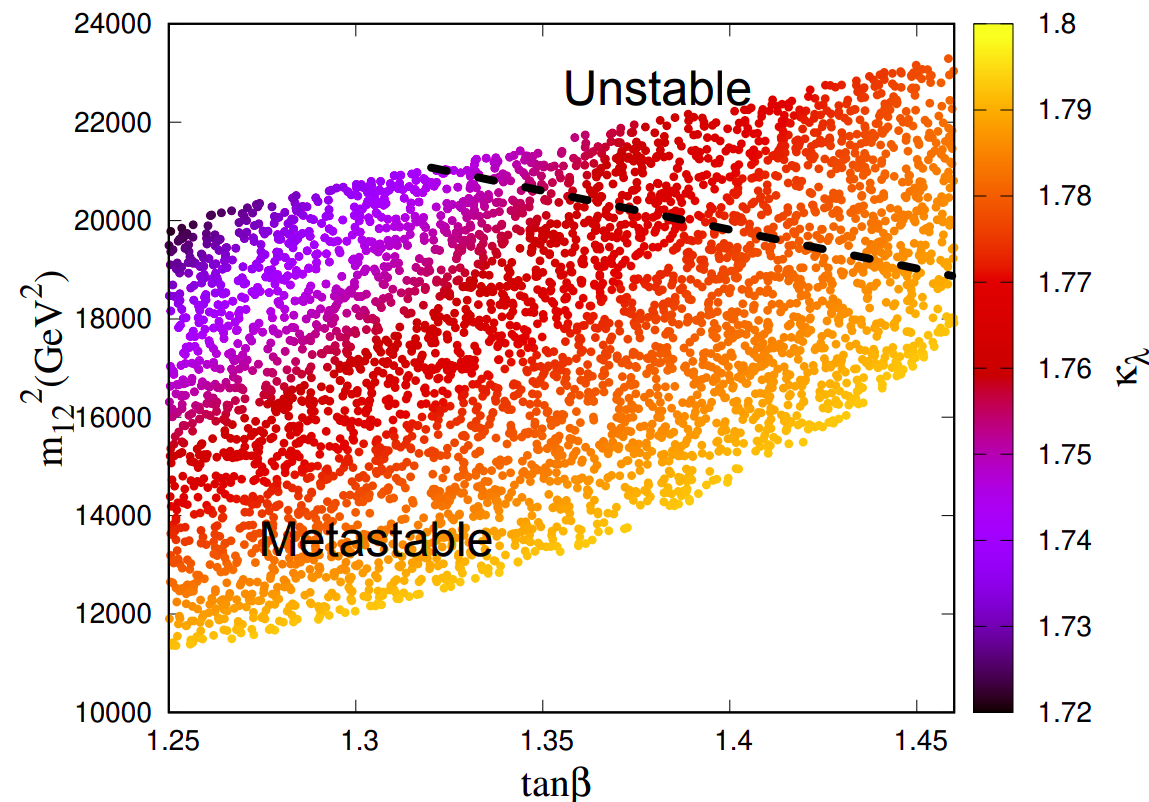} 
    \end{tabular}
\caption{N2HDM (top panel) and 2HDMS (bottom panel) with $\kappa_{\lambda}=1.4$ in $\tan\beta-m_{12}^2$ plane. $\Delta \kappa_\lambda /\kappa_\lambda \lesssim$ 10\% (left) and 30\% (right) for 2HDMS case. }
\label{trilinear_1p4}
\end{figure}

\begin{figure}
\centering
    \begin{tabular}{c}
        \textbf{N2HDM} \\
        \includegraphics[width=7.5cm,height=5cm]{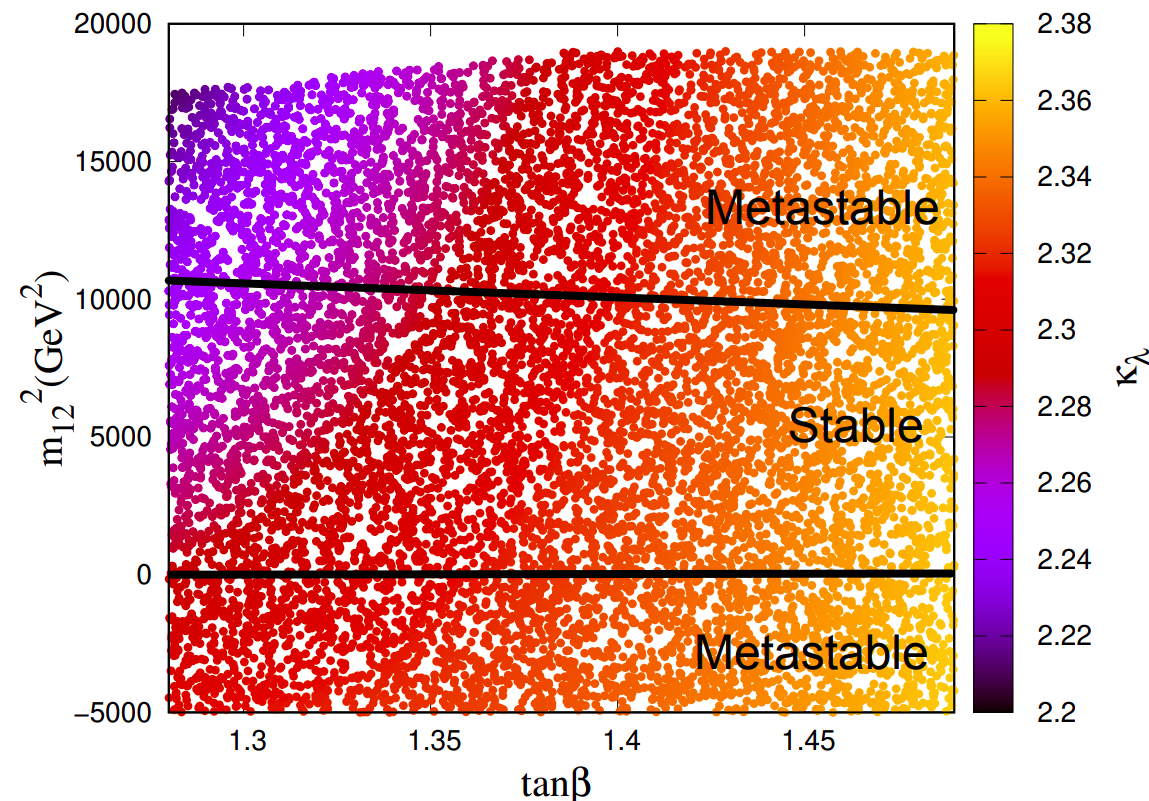}
        \end{tabular}
    \begin{tabular}{c c}
        \textbf{2HDMS} & \textbf{2HDMS} \\
        \includegraphics[width=7.5cm,height=5cm]{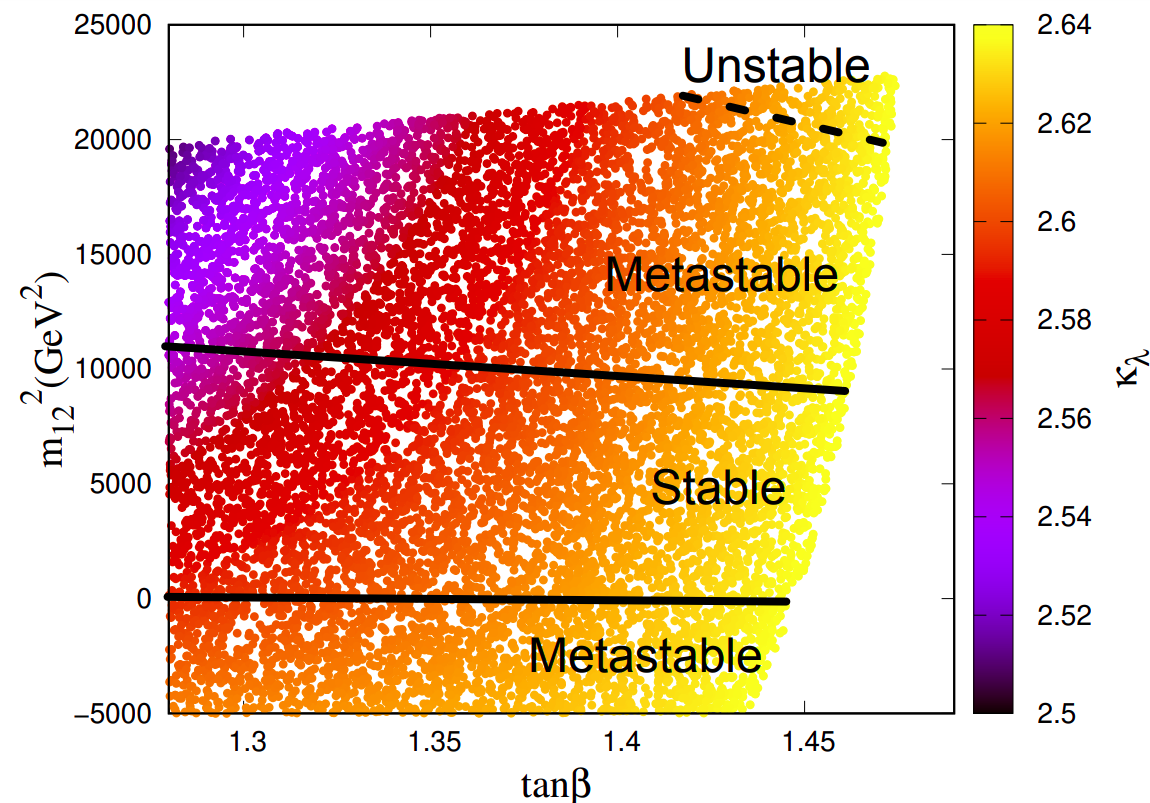} &
        \includegraphics[width=7.5cm,height=5cm]{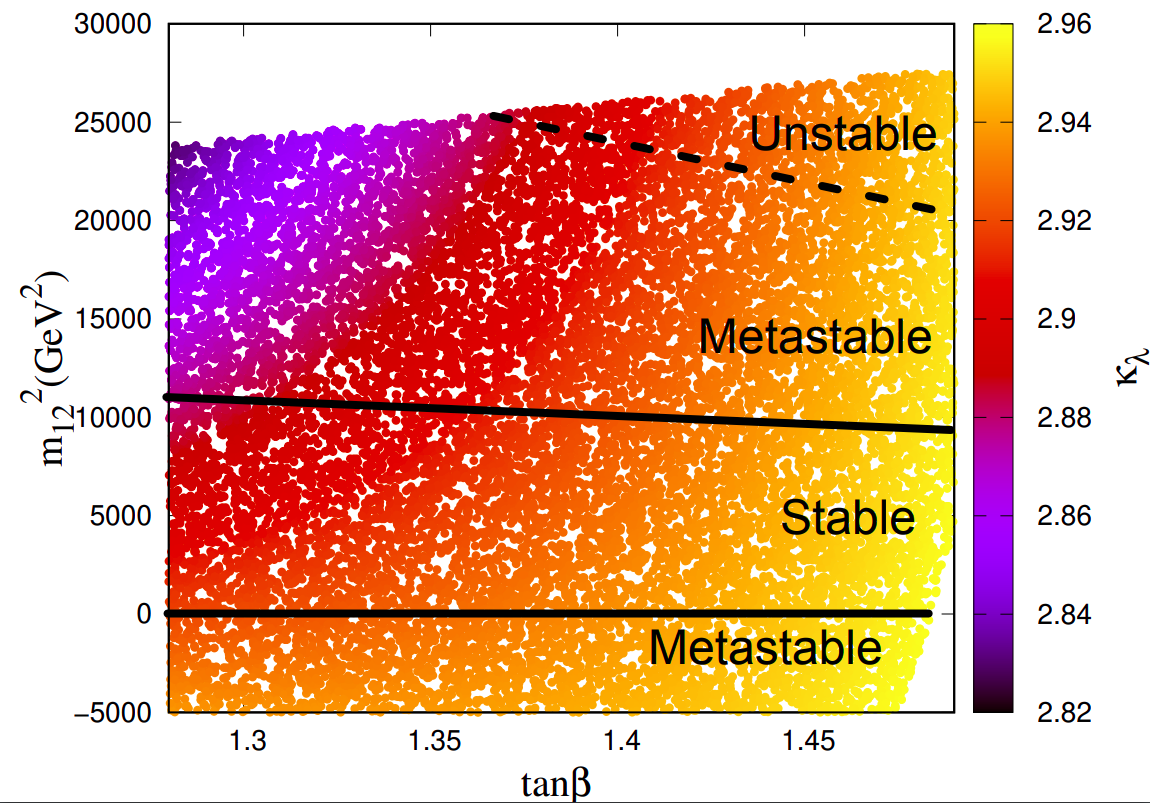} 
    \end{tabular}
\caption{N2HDM (top panel) and 2HDMS (bottom panel) with $\kappa_{\lambda}=2.3$ in $\tan\beta-m_{12}^2$ plane. $\Delta \kappa_\lambda /\kappa_\lambda \lesssim$ 10\% (left) and 30\% (right) for 2HDMS case. }
\label{trilinear_2p3}
\end{figure}

\noindent
In the preceding section, we focused on the signal strengths of the scalars in N2HDM and 2HDMS, which involve Yukawa and gauge boson couplings of the scalars. However, we have made no assumption on the self couplings, more specifically, trilinear couplings of the 125 GeV Higgs. We will discuss the impact of trilinear coupling of the 125 GeV Higgs boson denoted as $\lambda_{hhh}$. 

The measurement of $\lambda_{hhh}$ is absolutely crucial 
in understanding the nature of the scalar potential away 
from the EW vacuum. On the one hand, 
the measurement of 
trilinear coupling can probe the Electroweak phase 
transition. On the other hand, it can also probe the
deviation of the scalar sector from the SM. Therefore, the 
precise measurement of trilinear Higgs  coupling at the LHC 
as well as at the future colliders is of significant importance. The bound on 
$\lambda_{hhh}$ is usually given in terms of its 
relative strength with respect to its SM prediction, i.e 
$\kappa_{\lambda} = \lambda_{hhh}/\lambda_{hhh}^{SM}$. The 
most stringent bound on $\kappa_{\lambda}$ comes from the
LHC, $-0.4 < \kappa_{\lambda} < 6.3$ at 95\%
C.L~\cite{ATLAS:2022jtk}. More precise determination of $\kappa_{\lambda}$ is expected at future $e^+e^-$ colliders with $\sqrt{s} \geq 500$ GeV~\cite{deBlas:2019rxi} 


The SM prediction of $\lambda_{hhh}$ is 3$m_h^2/v$. However, this number can significantly change once extra scalars are introduced in a model. This is indeed the case for the N2HDM as well as the 2HDMS. We have chosen two specific values of $\kappa_{\lambda}$, that are allowed by the latest LHC bound~\cite{ATLAS:2022jtk}. We choose one case with $\kappa_{\lambda} \approx 2.3$. This value of $\kappa_{\lambda}$ represents the most challenging scenario for the measurement of $\kappa_{\lambda}$ through di-Higgs production at the LHC due to destructive interference between the diagrams involving the tri-Higgs couplings and the one without it, resulting in very small cross-section. We also consider another value for $\kappa_{\lambda} \approx 1.4$ for comparison.

In Figure~\ref{trilinear_1p4}, we show the N2HDM and 2HDMS parameter space consistent with $\kappa_{\lambda} \approx 1.4$. The masses and mixing angles in the N2HDM are consistent with the stable benchmark BP1 in Table~\ref{bp1_massbasis}. $m_{12}^2$ has been varied freely while $\tan\beta$ has been varied only within a small range compared to the benchmark value presented in Table~\ref{bp1_massbasis}. In case of the 2HDMS as well, the masses and three mixing angles($\alpha_{1,2,3})$ are kept at the similar values as in Table~\ref{bp1_massbasis}. But the other free parameters namely
$\alpha_{4,5,6}$ and the singlet vev $v_p$ are varied.
The variation of $m_{12}^2$ is again free and $\tan\beta$ is varied within a small range. We show the region allowed in the 2HDMS in $\tan\beta-m_{12}^2$ plane for  $\Delta \kappa_\lambda /\kappa_\lambda \lsim$ 10\% (bottom left) and 30\%(bottom right). The value of  $\kappa_\lambda$ is presented as the color axis.

One can clearly see that there is some correlation between the the value of $\kappa_{\lambda}$ and stability of a region. When we allow 30\% uncertainty in the measurement of $\kappa_{\lambda}$(bottom right plot), there can appear some regions with unstable vacuum and those regions should be excluded. On the other hand a more precise measurement of $\kappa_{\lambda}$, with an uncertainty of 10\% will determine the parameter space as metastable for this benchmark. On the other hand, in case of the N2HDM, the theoretical constraints namely, boundedness from below and perturbative unitarity, turn out to be stronger constraints (due to the absence of the additional free parameter unlike the 2HDMS case) compared to the constraints from $\kappa_{\lambda}$ measurement. Therefore the region shown in Figure~\ref{trilinear_1p4}(top) is consistent with $\Delta \kappa_\lambda /\kappa_\lambda \lsim$ 10\% as well as $\Delta \kappa_\lambda /\kappa_\lambda \lsim$ 30\%.

In Figure~\ref{trilinear_2p3}, we present the allowed parameter space for $\kappa_{\lambda} \approx 2.3$. Here too, the masses and mixing angles ($\alpha_{1,2,3}$) are consistent with BP1 in Table~\ref{bp1_massbasis}. In 2HDMS, $\alpha_{4,5,6}$ and the singlet vev $v_p$ are varied. Figure~\ref{trilinear_2p3} (top) shows the allowed parameter space for the N2HDM.  In Figure~\ref{trilinear_2p3}(left) and (right) the 2HDMS parameter space regions consistent with $\Delta \kappa_\lambda /\kappa_\lambda \lsim$ 10\% and $\Delta \kappa_\lambda /\kappa_\lambda \lsim$ 30\% are shown respectively. It is clear that this scenario with $\kappa_{\lambda} \approx 2.3$ is less stringent compared to $\kappa_{\lambda} \approx 1.4$ case, from the point of view of the $\kappa_\lambda$ measurement.

We mention here that, the self-coupling of the scalars directly involve the quartic couplings in the potential and therefore, are more sensitive to the vacuum stability compared to the Yukawa or gauge boson couplings where the impact of the quartic couplings is only indirect. 

\subsubsection{Probing $m_{12}^2$ via trilinear coupling}

It is well-known that the soft $Z_2$-breaking mass parameter $m_{12}^2$ is a parameter of the 2HDM as well as of models with extended scalars such as the N2HDM and the 2HDMS, which can not be directly determined from any specific observations. However, through the mass relations, pertinent to specific models, there is a strong correlation between the parameter $m_{12}^2$ and the quartic couplings of the scalar potential. Therefore, once the masses of the scalars are measured with the desired precision, one can get a strong constraints for the parameter $m_{12}^2$ from the trilinear coupling measurement. As it can be seen from Figures~\ref{trilinear_1p4} and ~\ref{trilinear_2p3}, that in our two models, namely the N2HDM and the 2HDMS, the limits on $m_{12}^2$ changes significantly and that can act strongly as a discriminant between the two models. Furthermore, one can expect a better discrimination if a more precise measurement of the trilinear coupling of the Higgs or equivalently $\kappa_{\lambda}$ is achievable. 

Having discussed the importance of the trilinear coupling measurement in distinguishing the two models and in extracting information on vacuum stability in the two, we should mention one important caveat. As we have discussed earlier, all our analyses so far are done at the tree-level. However, in~\cite{Bahl:2023eau,Aiko:2023nqj} it has been pointed out that the higher-order corrections to the diagrams leading to the Higgs self-coupling determination can change the results significantly. Therefore, while constraining the parameter space from future trilinear Higgs coupling measurement can be extremely effective, one should include the higher-order contributions in order to get the correct interpretation. 

\section{Exploring the difference between N2HDM and dark 2HDMS : ${\cal N}_s$-type vacua}
\label{sec6}

\begin{figure}[!hptb]
	\centering
	\includegraphics[width=7.5cm,height=6.5cm]{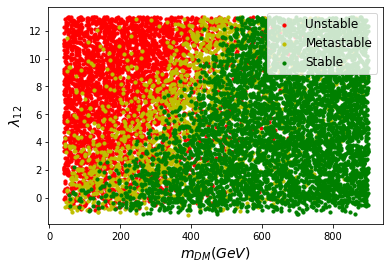}
        \includegraphics[width=7.5cm,height=6.5cm]{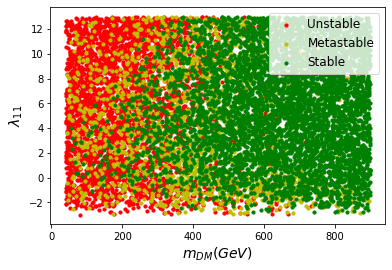} \\
        \includegraphics[width=7.5cm,height=6.5cm]{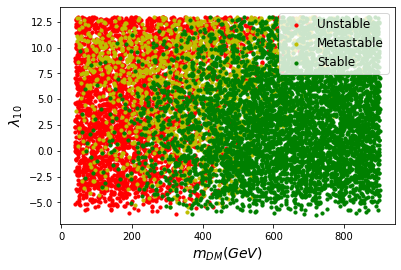}
	\caption{Panels:(top left) $m_{\text{DM}}-\lambda_{11}$, (top right) $m_{\text{DM}}-\lambda_{12}$, (bottom) $m_{\text{DM}}-\lambda_{10}$ in the 2HDMS for BP1. No DM constraints have been applied yet.}
	\label{2hdms_ns_bp1}
\end{figure}

\begin{figure}[!hptb]
	\centering
	\includegraphics[width=7.5cm,height=6.5cm]{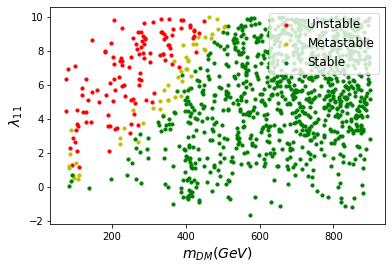}
        \includegraphics[width=7.5cm,height=6.5cm]{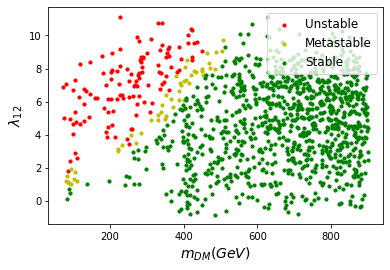} \\
        \includegraphics[width=7.5cm,height=6.5cm]{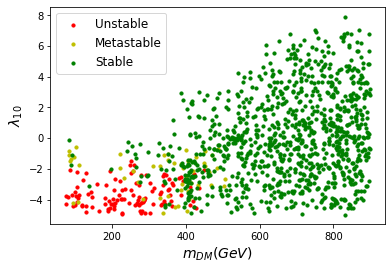}
	\caption{Panels: (top left) $m_{\text{DM}}-\lambda_{11}$, (top right) $m_{\text{DM}}-\lambda_{12}$, (bottom) $m_{\text{DM}}-\lambda_{10}$ in the 2HDMS for BP1. All points in these three plots satisfy the relic under-abundance, direct search and indirect search constraints.}
	\label{2hdms_ns_bp1_dm}
\end{figure}

\noindent
We now proceed to explore the difference between the two models for ${\cal N}_s$-type vacua. In those cases, only the real component of the singlet gets a vev within the 2HDMS framework. Since the imaginary component does not acquire a vev, it does not mix with other scalar states and is absolutely stable under the imposed discrete $Z_2'$ symmetry on the scalar potential. Therefore, this scenario can give rise to a viable DM candidate in the 2HDMS and is of genuine interest.  However, one should note that, in the N2HDM case, this type of scenario (${\cal N}_s$-vacuum) does not lead to a DM sector at all. That is an important distinction between the two models as well. 

Since in the 2HDMS, the imaginary component of the complex singlet does not acquire a vev in this case, there is no extra mixing in the scalar sector, compared to the N2HDM. Consequently, the visible sector will remain exactly the same in both the N2HDM and the 2HDMS. On the other hand, the additional parameters of the 2HDMS, namely $m_S'^2, \lambda_9$, $\lambda_{10}$, $\lambda_{11}$ and $\lambda_{12}$, remain completely decoupled from the visible sector, i.e they do not affect the observable spectrum of the visible sector. However, they will have an impact on the invisible branching fractions of the scalars. 

Let us consider BP1 of the N2HDM, which has been checked to be a stable point. Moving to the 2HDMS, we keep the parameters of the subspace of the 2HDMS fixed, which correspond to the N2HDM limit(BP1), and vary freely the extra parameters. In Figure~\ref{2hdms_ns_bp1}, we show the parameter space in the plane of the DM mass and the DM portal couplings ($\lambda_{10}, \lambda_{11}$ and $\lambda_{12}$). One can see that large unstable regions can appear, especially for small DM masses and large portal couplings. The relation between the mass of the DM candidate and the portal couplings can be written as follows:

\begin{equation}
m_{DM}^2 = m_S'^2 + \frac{1}{4}\lambda_{10} v_s^2 + \frac{1}{2}(\lambda_{11} v_1^2 + \lambda_{12}v_2^2). 
\label{dmmass}
\end{equation}

\noindent
It is worth mentioning that all the plots in Figure~\ref{2hdms_ns_bp1} are marginalized, i.e while the region is shown in the plane of two parameters, the other parameters are varied independently within the perturbative limit. No other constraints are applied on the dark sector parameter space in Figure~\ref{2hdms_ns_bp1}.

However, both DM mass as well as DM portal couplings receive substantial constraints from the observed relic density $\Omega h^2 = 0.12 \pm 0012$ from the Planck experiment~\cite{Planck:2018vyg}, the direct detection bounds from latest LUX-ZEPLIN experiments~\cite{LZ:2022lsv} and from indirect detection bounds from the Fermi-LAT experiments~\cite{Gammaldi:2021zdm}. We have considered the parameter space which gives rise to the observed relic density within $2\sigma$-uncertainty region. In addition, we have considered only the DM mass range $m_{DM} > 62.5$ GeV, in order to avoid the decay of the 125 GeV Higgs into invisible final states, adhering to strong constraint (BR(Higgs $\rightarrow invisible \lsim 10\%$)) from the 95\% upper limit on invisible branching ratio of the 125 GeV Higgs~\cite{ATLAS:2023tkt,CMS:2023sdw}. It is in principle possible to achieve finetuned regions of parameters below the 
 mass range considered here, by suppressing the coupling of 125 GeV Higgs to DM pairs. But we do not consider such scenarios in this work. It is also worth mentioning that the large portal couplings and the low DM mass may lead to negative $m_S'^2$ and consequently inconsistencies with the DM acquiring a zero vev. We have checked for such regions in our parameter scan and excluded them. Interestingly, in our case, we have three portal couplings at our disposal and we can get freedom in this context with relative sign difference among the relevant couplings.

We explored, that once the aforementioned constraints are applied on the 2HDMS parameter space, namely on the DM mass and portal couplings, the difference in vacuum stability in the 2HDMS compared to the N2HDM, still persists.
In Figure~\ref{2hdms_ns_bp1_dm}, we present the allowed parameter space from all the DM observables mentioned above, in the planes of the DM mass and DM portal couplings. Remarkably one can see that even in the limited parameter space allowed by the DM constraints, one can have unstable points in the 2HDMS, while the N2HDM limit corresponds to an absolutely stable point.

\section{Conclusion}
\label{sec7}

It is well-known that the stability of the EW vacuum can be jeopardized when the SM is extended with additional scalars. This is the case even at the leading order and zero temperature. In this work, we explore the vacuum stability of two extended scalar models, namely the N2HDM, where the scalar sector contains two scalar doublets and a real scalar singlet and the 2HDMS where there are two scalar doublets and a complex singlet. The two models are particularly interesting because both of them can accommodate the recently observed excess around 95 GeV, at the CMS and later at ATLAS in the $\gamma\gamma$ final state, with the observed signal strength. The 95 GeV excess can be accommodated in N2HDM when the real scalar singlet acquires a vev (termed as ${\cal N}_s$ type minima in this work). In 2HDMS, on the other hand, the 95 GeV excess can be explained when either (${\cal N}_s$/${\cal N}_p$ minima) or both components of the complex singlet acquire a vev (${\cal N}_{sp}$ minima). One more important feature of the 2HDMS is that, when one of the components of the complex singlet acquires vev, the other component can act as a dark matter component. Therefore, it is possible to explain the 95 GeV excess and also provide a viable DM candidate simultaneously in the 2HDMS (termed as dark 2HDMS in this work), which is not the case in the N2HDM. There, only one of the two scenarios can take place at a time.  

In order to achieve that we make a comparison between the vacuum structures of the two models. For that, we studied the absolute stability conditions in both models analytically for all possible types of EW vacua. Next we take up the numerical analysis and choose a specific benchmark in the N2HDM which can accommodate the 95 GeV excess and is absolutely stable. We compare the vacuum stability of this parameter point in the 2HDMS with both ${\cal N}_s$- and ${\cal N}_{sp}$-type minima. The comparison in the ${\cal N}_s$ case (dark 2HDMS) is simpler, since in this case the additional degree of freedom of 2HDMS completely decouple from the scalar scalar of the N2HDM owing to the zero vev condition. Therefore the visible sector in both models will lead to an exact same observable spectrum. However the phenomenology involving the additional scalar degrees of freedom will have impact on the DM sector in the 2HDMS and also on the vacuum stability of the parameter points. We have presented the interrelation between the two.  

On the other hand, the comparison of the N2HDM (${\cal N}_s$) minima with the 2HDMS (${\cal N}_{sp}$) is more complicated, since there the observable spectrum in both models differ from each other significantly due to extra mixing with the additional degree of 2HDMS. Therefore, we have first performed here a numerical study of the vacuum (in)stabilities with the additional parameters of the 2HDMS varying freely and examined their impact general. We have also considered the observable spectrum and demanded that the masses and observable signal strengths of all the scalars are within the 10-15\% of each other in the case of the two models. Thereafter, we have explored the vacuum stabilities in the allowed parameter space for both the two models. It has been shown that even with the little freedom on the additional parameter space in the 2HDMS, it is possible to have different vacuum structures in the 2HDMS compared to the N2HDM. It has also been pointed out that although the Yukawa and gauge boson coupling measurements put constraints on vacuum stability, their impact is rather indirect. On the other hand, the self-couplings of the scalars, e.g trilinear coupling of 125 GeV Higgs turns out to be a better probe of the vacuum structure, since it is directly connected to the quartic couplings as well as to the soft $Z_2$-breaking mass term $m_{12}^2$. We have also shown that, it is possible to probe the parameter $m_{12}^2$ (which is otherwise inaccessible), via the measurement of the trilinear Higgs coupling, which can furthermore distinguish between the two models.

\section*{Acknowledgements}
JL and GMP acknowledge support by the Deutsche Forschungsgemeinschaft (DFG, German Research Foundation) under Germany's Excellence Strategy EXC 2121 "Quantum Universe"- 390833306. JL and GMP would like to thank Georg Weiglein, Thomas Biek{\"o}tter and Fabio Campello for useful discussions and valuable suggestions. JL would like to thank Fabio Campello for providing important inputs regarding the code EVADE.

\begin{appendix}

\section{Comparison with two real scalars}
\label{tworealscalar}

Once we write the terms involving $\Phi_S$ in the scalar potential in Eq.~(\ref{eq:cspot}) in terms of the real and imaginary component of $\Phi_S$, namely $S$ and $P$, we arrive at Equation~\ref{2hdms_lag}, if we assume all the coefficients in \ref{eq:cspot} are real. Under this assumption terms such as $S^3 P$, $P^3 S$ or $SP$ cancel due to the requirement of hermiticity of the Lagrangian. Furthermore, the scalar potential in Equation~\ref{2hdms_lag} is analogous to a scalar potential involving two scalars $S$ and $P$, but necessarily charged under two different $Z_2'$-symmetries. 

On the other hand, if one considers a scalar potential involving two real scalars $S$ and $P$, both charged under the same $Z_2'$ symmetry, the aforementioned terms $S^3 P$, $P^3 S$ or $SP$ can occur, since they are allowed by the symmetry. In the complex singlet scenario, however, these terms can appear only with complex coefficients.  

Therefore, a complex singlet extension of 2HDM and two real scalar extension of 2HDM can in principle be used interchangeably, but with the aforementioned small caveat in mind.

\section{Comparison with $Z_3$-symmetric 2HDMS}
\label{z32hdms_compare}

In~\cite{Heinemeyer:2021msz}, a 2HDMS has been studied as well. The difference between that model and the model considered in this work lies in the different symmetry structure. In \cite{Heinemeyer:2021msz}, the complex singlet is charged under $Z_3$ symmetry. The imposed $Z_3$ symmetry in that case, is motivated by the supersymmetric version NMSSM~\cite{Domingo:2018uim,Choi:2019yrv,Biekotter:2021qbc} 
The scalar potential pertaining to a $Z_3$-symmetric 2HDMS, is given as follows.

\begin{equation}
       \label{z32hdms}
	\begin{split}		V=&m_{11}^2\Phi_1^\dagger\Phi_1+m_{22}^2\Phi_2^\dagger\Phi_2-(m_{12}^2\Phi_1^\dagger\Phi_2+\mathrm{h.c.})\\
		&+\frac{\lambda_1}{2}(\Phi_1^\dagger\Phi_1)^2+\frac{\lambda_2}{2}(\Phi_2^\dagger\Phi_2)^2+\lambda_3(\Phi_1^\dagger\Phi_1)(\Phi_2^\dagger\Phi_2)+\lambda_4(\Phi_1^\dagger\Phi_2)(\Phi_2^\dagger\Phi_1)\\
		&+\Big[\frac{\lambda_5}{2}(\Phi_1^\dagger\Phi_2)^2+\lambda_6(\Phi_1^\dagger\Phi_1)(\Phi_1^\dagger\Phi_2)+\lambda_7(\Phi_2^\dagger\Phi_2)(\Phi_1^\dagger\Phi_2)+\mathrm{h.c.}\Big]\\
		&+(\xi \Phi_S+\mathrm{h.c.})+m_S^2 \Phi_S^{*} \Phi_S+(\frac{{m'_S}^2}{2}\Phi_S^2+\mathrm{h.c.})\\
		&+\left(\frac{\mu_{S1}}{6}\Phi_S^3+\frac{\mu_{S2}}{2}\Phi_S \Phi_S^{*} \Phi_S+\mathrm{h.c.}\right)+\left(\frac{\lambda''_1}{24}\Phi_S^4+\frac{\lambda''_2}{6}\Phi_S^2 \Phi_S^{*} \Phi_S+\mathrm{h.c.}\right)+\frac{\lambda''_3}{4}(\Phi_S^{*} \Phi_S)^2\\
		&+\Big[\Phi_S(\mu_{11}\Phi_1^\dagger\Phi_1+\mu_{22}\Phi_2^\dagger\Phi_2+\mu_{12}\Phi_1^\dagger\Phi_2+\mu_{21}\Phi_2^\dagger\Phi_1)+\mathrm{h.c.}\Big]\\
		&+\Phi_S^{*} \Phi_S\Big[\lambda'_1\Phi_1^\dagger\Phi_1+\lambda'_2\Phi_2^\dagger\Phi_2+\lambda'_3\Phi_1^\dagger\Phi_2+\mathrm{h.c.}\Big]\\
		&+\Big[\Phi_S^2(\lambda'_4\Phi_1^\dagger\Phi_1+\lambda'_5\Phi_2^\dagger\Phi_2+\lambda'_6\Phi_1^\dagger\Phi_2+\lambda'_7\Phi_2^\dagger\Phi_1)+\mathrm{h.c.}\Big]
	\end{split}
\end{equation}

\noindent
where $\Phi_S = S + iP$ denotes the complex scalar field. Under $Z_3$ symmetry, the fields transform in the following way.

\begin{align}
    {\rm 2HDMS~}\quad Z_3 &~:~ 
    	\begin{pmatrix} \Phi_1\\ \Phi_2\\ \Phi_S \end{pmatrix} \to 
    	\begin{pmatrix} 
    	1&	&	\\	&	e^{i2\pi/3}&	\\	&	&	e^{-i2\pi/3}
	   \end{pmatrix}\,
	   \begin{pmatrix} \Phi_1\\ \Phi_2\\ \Phi_S
	\end{pmatrix}
\end{align}

\noindent
One can check, that the terms involving $\Phi_S\Phi_1^\dagger\Phi_2$ or $\Phi_S\Phi_2^\dagger\Phi_1$ are responsible for mixing between the pseudoscalar degree of freedom $P$ and the pseudoscalar field of the 2HDM, namely $A$, giving rise to two pseudoscalar mass eigenstates $a_1, a_2$. The scalar states of the 2HDM mix with $S$ and give rise to three scalar mass eigenstates $h_1,h_2,h_3$. It can also be verified that the aforesaid pseudoscalar mixing angle depends on the couplings  $\mu_{12}$ and $\mu_{21}$. 

On the other hand, the scalar potential of 2HDMS with $Z_2'$, as considered in the present work, can be written as in~\cite{Dutta:2023cig}:

\begin{align}
\label{eq:cspot}
   V &= m^2_{11} \Phi_1^{\dagger}\Phi_1 + m^2_{22} \Phi_2^{\dagger}\Phi_2-(m^2_{12} \Phi_1^{\dagger}\Phi_2+h.c)  \nonumber \\ 
   & \quad + \frac{\lambda_1}{2} (\Phi_1^{\dagger}\Phi_1)^2 +  \frac{\lambda_2}{2} (\Phi_2^{\dagger}\Phi_2)^2 + \lambda_3 (\Phi_1^{\dagger}\Phi_1) (\Phi_2^{\dagger}\Phi_2) \nonumber \\
   & \quad + \lambda_4 (\Phi_1^{\dagger}\Phi_2) (\Phi_2^{\dagger}\Phi_1) + [\frac{\lambda_5}{2} (\Phi_1^{\dagger}\Phi_2)^2+h.c] \nonumber \\
   & \quad + m^2_S \Phi_S^{*}\Phi_S + (\frac{m^{\prime}_S}{2} \Phi_S^2 + h.c) \nonumber \\
   & \quad  + (\frac{\lambda^{\prime\prime}_1}{24}\Phi_S^4 +h.c)+(\frac{\lambda_2^{\prime\prime}}{6}(\Phi_S^2 \Phi_S^{*}\Phi_S) +h.c)+\frac{\lambda_3^{\prime\prime}}{4} (\Phi_S^{*}\Phi_S)^2 \nonumber \\
   & \quad + \Phi_S^{*}\Phi_S[\lambda_1^{\prime} \Phi_1^{\dagger}\Phi_1 + \lambda_2^{\prime}\Phi_2^{\dagger}\Phi_2] +  [\Phi_S^2 (\lambda^{\prime}_4 \Phi_1^{\dagger}\Phi_1+\lambda_5^{\prime}\Phi_2^{\dagger}\Phi_2) +h.c.].
\end{align}

\noindent
Comparing Eqs.~(\ref{z32hdms}) and (\ref{eq:cspot}), we can see that in the $Z_2'$-symmetric scenario, the terms $\Phi_S\Phi_1^\dagger\Phi_2$ are absent, since they are forbidden by the symmetry. Because of the absence of these particular terms, there is no mixing between $P$ and the pseudoscalar $A$ of the 2HDM. On the contrary, when $P$ acquires a vev, $S$ and $P$ both mix with the scalar states of the 2HDM and give rise to four scalars $h_1,..h_4$, while $A$ still remains the only pseudoscalar mass eigenstate. If $v_p = 0$, the scalar states of 2HDM mix with $S$ giving rise to three scalar states $h_1,..h_3$ and $P$ does not mix with any other state and behaves like a stable DM candidate.  

Furthermore, it is noteworthy that if in the $Z_3$-symmetric 2HDMS, the pseudoscalar $P$ acquires a vev, it introduces to spontaneous CP-violation, i.e. mixing between scalar and pseudoscalar states. On the other hand, in $Z_2'$-symmetric 2HDMS, the vev of $P$ does not lead to spontaneous CP-violation, rather it leads to mixing of $P$ with other scalar states, as discussed earlier. 

One crucial difference between the two models is, since in the $Z_2'$-symmetric case, both N2HDM and 2HDMS are charged under the same symmetry, i.e. $Z_2'$-symmetry, it is possible to reach the N2HDM limit in the 2HDMS. But in the $Z_3$-symmetric case, since N2HDM and 2HDMS are charged under two different discrete $Z_2$-symmetries, it is not possible to attain the N2HDM as a limiting scenario of the $Z_3$-symmetric 2HDMS.

\section{Detailed calculation of (in)stability condition}
\label{calculation}

We have derived the conditions for vacuum stability in Section~\ref{sec3}. Here we present a detailed calculation for one of the cases considered. We consider the scenario of charge breaking of ${\cal CB}$-type of the ${\cal N}_s$-type EW vacuum in the 2HDMS.  

In terms of the bilinears($x_1, x_2, x_3, x_4, x_5, x_6$) the scalar potential can be written as 
\begin{equation}
V = A^T X + \frac{1}{2}X^T B X,
\label{potential}
\end{equation}

where

\begin{equation}
X = \left(\begin{array}{c} x_1 \\ x_2 \\ x_3 \\ x_4 \\ x_5 \\ x_6
  \end{array}\right)\,,\quad
A = \left(\begin{array}{c} m_{11}^2 \\ m_{22}^2 \\ -2 m_{12}^2 \\ 0 \\ m^2_S \\ m_{S'}^2
  \end{array}\right)\,,\quad
B = \left(\begin{array}{cccccc} \lambda_1 & \lambda_3 & 0                        & 0                        & \lambda_7 & \lambda_{11}\\
    \lambda_3         & \lambda_2 & 0                        & 0                        & \lambda_8 & \lambda_{12} \\
    0                 & 0         & 2(\lambda_4 + \lambda_5) & 0                        & 0   & 0      \\
    0                 & 0         & 0                        & 2(\lambda_4 - \lambda_5) & 0    & 0      \\
    \lambda_7         & \lambda_8 & 0                        & 0                        & \lambda_6   & \lambda_{10} \\ \lambda_{11} & \lambda_{12} & 0 & 0 & \lambda_{10} & 
  \lambda_9 \end{array}\right) \, .
\label{eq:def}
\end{equation}

In the ${\cal N}_s$-vacuum, we have
\begin{equation}
\braket{\Phi_1}_0=\frac{1}{\sqrt{2}}\left(\begin{array}{c} 0 \\
\,v_1
\end{array}\right)\,, \ \ \
\braket{\Phi_2}_0=\frac{1}{\sqrt{2}}\left(\begin{array}{c} 0 \\
\,v_2
\end{array}\right)\,, \ \ \
\braket{S}_0=v_s, \ \ \ \braket{P}_0=0
\end{equation}

with 

\begin{equation}
X_{{\cal N}_s}  = \langle X\rangle_{{\cal N}_s} = \frac{1}{2}\,
\left(\begin{array}{c} v_1^2 \\ v_2^2 \\ v_1 v_2 \\ 0 \\ v_s^2 \\ 0 \end{array}\right)\
\,,\quad
V^\prime_{{\cal N}_s}  =  A\,+\,B\,X_{{\cal N}_s} =
\left(\begin{array}{c} \frac{v_2^2}{v^2} m^2_{H^\pm} \\ \frac{v_1^2}{v^2} m^2_{H^\pm}
    \\ - \frac{2 v_1 v_2}{v^2} m^2_{H^\pm}
    \\ 0  \\ 0 \\ m_p^2\end{array}\right)\, ,
\label{eq:xvn1}
\end{equation}
where $v^2 = v_1^2 + v_2^2$. The entries of $V^\prime_{{\cal N}_s}$ can be obtained using the
minimization conditions and eigenvalues of the scalar mass
matrices, where $m^2_{H^\pm}$ is the squared charged scalar mass at this
stationary point and $m_p^2$ the squared mass of the $P$ field. These are
given by
\begin{equation}
m^2_{H^\pm} = m^2_{12}\,\frac{v^2}{v_1 v_2} \,-\,\frac{1}{2}(\lambda_4 + \lambda_5)\, v^2
\,,\quad
m_p^2 = m_{S'}^2\,+\,\frac{1}{4} \lambda_{10} v_s^2\,+\,\frac{1}{2} (\lambda_{11} v_1^2\,+\,\lambda_{12} v_2^2)\,.
\end{equation}

In the ${\cal CB}$-vacuum

\begin{equation}
\braket{\Phi_1}_0=\frac{1}{\sqrt{2}}\left(\begin{array}{c} 0 \\
\,c_1
\end{array}\right)\,, \ \ \
\braket{\Phi_2}_0=\frac{1}{\sqrt{2}}\left(\begin{array}{c} c_2 \\
\,c_3
\end{array}\right)\,, \ \ \
\braket{S}_0=0,\ \ \ \braket{P}_0=0.
\end{equation}

\begin{equation}
        X_{\cal CB}  = \langle X\rangle_{\cal CB} = \frac{1}{2}\,
        \left(\begin{array}{c} c_1^2 \\ c_2^2 + c_3^2 \\ c_1 c_3 \\ 0 \\ 0 \\ 0\end{array}\right)\
        \,,\quad
        V^\prime_{\cal CB}  =  A\,+\,B\,X_{\cal CB} =
        \left(\begin{array}{c} 0 \\ 0 \\ 0 \\ 0 \\ m^2_{S1} \\ m^2_{S2} \end{array}\right) \,,
        \label{eq:xvcb1}
        \end{equation}
        where $m^2_{S1} = m_S^2 + \lambda_7 c_1^2/2 + \lambda_8 (c_2^2 +
          c_3^2)/2$, and
      $m^2_{S2} = m_{S'}^2 + \lambda_{11} c_1^2/2 + \lambda_{12} (c_2^2 +
          c_3^2)/2$
        are the squared scalar masses at the ${\cal CB}$-stationary point.
        Here too, the entries of $V^\prime_{\cal CB}$ can be obtained by ${\cal CB}$ minimization conditions.

The internal product of the vectors $X_{\cal CB}$ and $V^\prime_{{\cal N}_s}$ yields
\begin{equation}
X_{\cal CB}^T\,V^\prime_{{\cal N}_s} = \frac{m^2_{H^\pm}}{2 v^2}\,\left[(v_2 c_1 - v_1 c_3)^2 + v_1^2 c_2^2\right],
\label{eq:xcb1vn1}
\end{equation}
which may also be written as
\begin{equation}
X_{\cal CB}^T\,V^\prime_{{\cal N}_s}\;=\;  X_{\cal CB}^T \,\left(A \,+\,B\,X_{{\cal N}_s}\right)\;=\;
X_{\cal CB}^T\,A\,+\,X_{\cal CB}^T\,B\,X_{{\cal N}_s}\,.
\label{eq:xcb1vn1_2}
\end{equation}

The potential at the stationary point(${\cal CB}$) can be written as

\begin{equation}
\label{eq:Vmin}
    V_{\cal CB} = \frac{1}{2} A^T X_{\cal CB}.
\end{equation}

From Eq.~\ref{eq:Vmin}, one can write
\begin{equation}
X_{\cal CB}^T\,A\;=\; 2\,V_{\cal CB}\, ,
\end{equation}
and therefore, combining Eqs.~(\ref{eq:xcb1vn1}) and (\ref{eq:xcb1vn1_2}),

\begin{equation}
X_{\cal CB}^T\,B\,X_{{\cal N}_s}  =  \frac{m^2_{H^\pm}}{2 v^2}\,\left[(v_2 c_1 - v_1 c_3)^2 + v_1^2 c_2^2\right]
\,-\,2\,V_{\cal CB}\,.
\label{eq:xx1}
\end{equation}
We now perform similar operations on the vectors $X_{{\cal N}_s}$ and $V^\prime_{\cal CB}$, yielding
\begin{equation}
X_{{\cal N}_s}^T\,V^\prime_{\cal CB} = v_s^2 m_{S1}^2\,\Leftrightarrow\,X_{{\cal N}_s}^T\,A\,+\,X_{{\cal N}_s}^T\,B\,X_{\cal CB} = v_s^2 m_{S1}^2\,.
\end{equation}
As before, the quantity $X_{{\cal N}_s}^T\,A$ is twice the value of the potential at the extremum ${\cal N}_s$. Therefore
\begin{equation}
X_{{\cal N}_s}^T\,B\,X_{\cal CB} = v_s^2 m_{S1}^2\,-\,2\,V_{{\cal N}_s}\,.
\label{eq:xx2}
\end{equation}
Since the matrix $B$ (defined in Eq.~(\ref{eq:def})) is symmetric, the left-hand sides of
Eqs.~(\ref{eq:xx1}) and~(\ref{eq:xx2}) are identical. It is then trivial to obtain the following
expression comparing the depth of the potential at both extrema,
\begin{equation}
V_{\cal CB} \,-\,V_{{\cal N}_s} = \frac{m^2_{H^\pm}}{4 v^2}\,\left[(v_2 c_1 - v_1 c_3)^2 + v_1^2 c_2^2\right]\,-\,v_s^2 m_{S1}^2.
\label{eq:vcb1n1}
\end{equation}

identical with Eq.~(\ref{eqn2}).

\end{appendix}

\bibliographystyle{JHEP}
\bibliography{ref}

\end{document}